\documentclass[aps,prd,twocolumn,tightenlines,showpacs,amsmath,amssymb,nofootinbib]{revtex4-1}

\usepackage{graphicx}
\usepackage{amsmath,amssymb}
\usepackage{bm}
\usepackage{epsfig}
\usepackage{color}              
\usepackage{comment}
\usepackage{hyperref}
\usepackage{array, float,tabularx}
\usepackage{lipsum}

\usepackage{amsmath}
\usepackage{slashed}

\def\be{\begin{equation}}
\def\ee{\end{equation}}
\def\ba{\begin{eqnarray}}
\def\ea{\end{eqnarray}}
\def\bal{\begin{align}}
\def\eal{\end{align}}
\def\bald{\begin{aligned}}
\def\eald{\end{aligned}}

\begin{document}

\title{Natural inflation, strong dynamics, and the role of generalized anomalies}

\date{\today}

\author{Mohamed M. Anber}
\email{manber@lclark.edu}
\author{Stephen Baker}
\email{sbaker@lclark.edu}
\affiliation{Department of Physics, Lewis \& Clark College, 
Portland, OR 97219, USA}

\begin{abstract}
We revisit models of natural inflation and show that the single-field effective theory described by the potential $V(a)\sim \cos\frac{a}{f}$ breaks down as the inflaton $a$ makes large-field excursions, even for values of $f$ smaller than the Planck scale. In order to remedy the problem,  we modify the potential in order to account for the heavy degrees of freedom (hadrons) that become intertwined with the light inflaton as the latter rolls down its potential.  By embedding the low energy degrees of freedom into an ultraviolet complete  gauge theory, we argue that the intertwining between the two scales  can be explained as the result of  a generalized mixed 't Hooft anomaly between the discrete chiral symmetry and background fractional fluxes in the baryon number, color, and flavor directions. Further, we study the multi-field inflation and show that it entertains rich dynamics. Inflating near the hilltop excites the hadrons and spoils the slow-roll parameters, in contradistinction with the expectations in the single-field inflation. Nevertheless, we identify a safe zone where inflation can proceed successfully.  We determine the  conditions under which the Universe inflates by at least $60$ e-foldings and  inflation leads to a power spectrum and tensor to scalar ratio that are consistent with the Cosmic Microwave Background data.
 \end{abstract}

\maketitle

\section{Introduction}

Inflationary models that make use of pseudo Nambu-Goldstone bosons are among the most compelling bottom-up particle physics approaches to inflation. They are said to be {\em natural} because the ratio of the inflaton mass to the Hubble parameter  is  protected from large  quantum corrections.  The first and most influential idea in this direction is to use   axions as inflatons \cite{Freese:1990rb,Adams:1992bn}. These pseudoscalar particles enjoy a continuous shift symmetry, $a\rightarrow a+\mbox{constant}$, that guards them against large loop corrections.  The shift symmetry is broken nonperturbatively, e.g., via anomalies, to a discrete shift symmetry, leading to a potential of the form:
\begin{eqnarray}
V(a)\sim\Lambda^4\left(1-\cos \frac{a}{f}\right)\,,
\label{axion effective V}
\end{eqnarray} 
where $f$ is the axion decay constant, or axion constant for brevity. The scale $\Lambda$ is the strong-coupling scale of the gauge group that causes the breaking of the shift symmetry. As one takes $\Lambda\rightarrow 0$ we restore the full shift symmetry, and hence, the flatness of the potential, which is related to the axion mass $m_a\approx \frac{\Lambda^2}{f}$, is natural in the sense of 't Hooft \cite{tHooft:1979rat}. 

The above potential has been extensively used in the literature for the purpose of studying axion inflation as well as other aspects of axion physics, see, e.g., \cite{Dimopoulos:2005ac,Anber:2006xt,Anber:2009ua,Adshead:2012kp,Remmen:2014mia}. Indeed,  $V(a)$ is a good effective field description at scales $\sim m_a $, which is much lower than $\Lambda$. However, as we shall argue in this letter, $V(a)$ breaks down as  $a$ makes a large-field excursion $a\sim f$, even for values of $f<M_P$, where $M_P$ is the reduced Planck mass\footnote{In this work we use the reduced Planck mass $M_P\equiv \frac{1}{\sqrt{8\pi G}}= 2.435\times10^{18}$ GeV, where $G$ is Newton's constant.}.
As $a$ traverses the field space, it becomes intertwined with heavy degrees of freedom that were integrated out in the first place and led to the potential (\ref{axion effective V}). This raises the question about the validity of $V(a)$ to study models of natural inflation. The main purpose of this letter is to address this concern and elucidate what really happens during axion inflation.

 Using chiral perturbation theory, we argue that the potential (\ref{axion effective V}) should be replaced with the {\em correct} effective potential that accounts for the heavy degrees of freedom:
\begin{eqnarray}
V(a,\sigma)\sim \Lambda^4\left(1-\cos\frac{\sigma}{\Lambda}\cos\frac{a}{f}\right)\,,
\label{low energy hadron and axion}
\end{eqnarray}
where $\sigma$ is the hadronic field\footnote{We abuse the language and use the word hadron to mean any strongly-coupled degree of freedom in the infrared, including glue balls.}.

On another but related topic, 't Hooft anomalies have been known to play a pivotal role in understanding the infrared (IR) physics of strongly coupled gauge theories  \cite{tHooft:1979rat}. In this regard, the  IR particle spectrum has to exactly match an 't Hooft anomaly that exists in the ultraviolet (UV). The failure to match an anomaly is a red flag that something is missing in the IR picture. Although this type of anomalies were known since the 80's, recently  it was realized that they can further be generalized and used to put extra constraints on the possible IR realization of the global symmetries of a given theory, see  \cite{Gaiotto:2014kfa,Gaiotto:2017yup,Anber:2019nze,Anber:2020gig} and references therein.  

As we discuss in this work, one can obtain $V(a)$ from a UV complete gauge theory and show that the IR effective  theory is missing important information about the matching of a new type of generalized 't Hooft anomaly that exists deep in the UV, known as the baryon number-color-flavor (BCF) anomaly \cite{Anber:2019nze,Anber:2020gig}.  In order to match the anomaly, we need to incorporate the hadronic degrees of freedom into $V(a)$. Here, however, one cannot make use of the chiral perturbation theory. The form of $V(a,\sigma)$ is motivated by lessons that have been learned over the past decade about {\em an adiabatic continuity} between the 4-D Yang-Mills theory and a circle compactified Yang-Mills theory with deformations, see \cite{Unsal:2008ch,Dunne:2016nmc}. The anomaly matching  conditions provide a non-perturbative and stronger  statement about the necessity of re-introducing the heavy degrees of freedom in the low-energy effective field theory as the axion makes large excursions in the field space.

Further, we use both numerical and analytical techniques to study the inflationary potential $V(a,\sigma)$. Unlike the single-field effective theory, where the axion can start rolling down the potential very close to the hilltop $a\approx \pi f$, starting near the maximum of $V(a,\sigma)$ spoils inflation for any value of $f$. The hadronic degrees of freedom kick the axion very hard towards a steep direction forcing inflation to end abruptly. However, if we start inflating in a {\em safe zone} at values of $a\approx \frac{\pi}{2}f$, then the axion rolls down slowly in a flat direction and inflation can be sustained. Whether we can have enough e-folds $\sim50-60$, which is the typical number required to solve the problems of the standard Big-Bang cosmology, depends on the value of $f$. Taking $f\lesssim M_P$ does not yield enough number of e-folds, which is dramatically different from inflating in $V(a)$, where one can achieve $\sim50-60$ e-folds of inflation even for values of $f\lesssim M_P$ by starting very close to the hilltop. This behavior is  a manifestation of the intertwining phenomenon between the axion and heavy degrees of freedom as the former makes a large-field excursion. We also argue that this complex dynamics is  ultimately attributed to the BCF anomaly, which is lurking deep in the IR and dictating the behavior of the system. 

Fortunately, we can achieve successful inflation that yields a large number of e-folds by taking $f> 9 M_P$. Although such large values of $f$ are in conflict with theories of quantum gravity, one can evade the problem, for example, by invoking several axions as in models of N-flation \cite{Dimopoulos:2005ac}.  Further, we study the curvature perturbations and show that the hadronic quantum fluctuations stay in the vacuum at the time the axion fluctuations exit the horizon.  This leads to values of the spectral tilt and tensor to scalar ratio that are compatible with the Cosmic Microwave Background (CMB) data \cite{Akrami:2018odb}.

This letter is organized as follows. In Section \ref{Theory and formulation} we discuss the reason behind the breakdown of the single-field potential and further describe a class of UV complete gauge theories coupled to fermions and a single or several complex Higgs fields and lead to a low energy effective theory  of axions. The UV theories are almost identical to the ones that were described in the original paper on natural inflation \cite{Adams:1992bn}, also see \cite{Shifman:1979if}. However, here we pay extra attention to the global symmetries, as was done in \cite{Anber:2020xfk}, since they play a pivotal role in identifying the new type of 't Hooft anomaly that constrains the IR dynamics, i.e., the BCF anomaly. Throughout this section we consider the Euclidean version of the theory in the background of a general manifold, but otherwise we turn off the dynamical gravity.  The Euclidean version is more convenient to work with since it is easier to identify all the global symmetries as well as their 't Hooft anomalies in a Euclidean setup. Then, we discuss what goes wrong with the low energy effective potential (\ref{axion effective V}). In order to remedy the effective theory, we use lessons  from Yang-Mills theory on a small circle with deformations. This class of theories respects the BCF anomaly, and hence, it provides clues about the form of the low energy effective potential (\ref{low energy hadron and axion}) that should replace (\ref{axion effective V}). In Section \ref{Dynamics of axion inflation} we turn on gravity and study the dynamics of the inflationary potential (\ref{low energy hadron and axion})  in the Friedmann-Robertson-Walker spacetime $ds^2=-dt^2+b^2(t)d\bm x^2$, where $b(t)$ is the scale factor and $t$ is the cosmic time. Finally, we study the curvature and tensor perturbations and compare against the results from the potential (\ref{axion effective V}). We conclude in Section \ref{Final Comments} with final comments. Many important and fine points are delegated to several footnotes, as we felt they might interrupt the flow of the main text.  

\section{Theory and formulation}
\label{Theory and formulation}

The plan of this section is as follows. We first discuss the breakdown of the single-field effective theory as the axion makes large excursions in the field space. For this purpose we use  chiral perturbation theory and show that one needs to take into account the hadronic degrees of freedom in order to cure the sick single-field theory. At this level one might be tempted to proceed right away and use the  two-field potential to study inflation. However, we take a long pause before doing that in order to show that there can be a deep reason, deeper than the rules of effective field theory, why one needs to  introduce the hadronic degrees of freedom into the effective potential. This reason is attributed to a new 't Hooft anomaly. However, in order to see how this works, one needs to embed the axion as well as hadrons in a UV complete gauge theory and traces the anomaly from the UV down to the IR. As it turns out, as we flow to the IR we lose control over the strong dynamics and it becomes unclear how to take into account the strongly-coupled degrees of freedom  and at the same time match the anomaly. Empowered with lessons we have learned over the past decade from a model of Yang-Mills theory compactified on a small circle, we propose a phenomenological model that: (1) takes into account the effects of the strong dynamics on axion and (2) matches the anomaly.

\subsection{Chiral Lagrangian} In this section we explain why the potential (\ref{axion effective V}) breaks down as the axion makes large excursions in the field space. Notice that this is true even for values of $f$ smaller then the Planck scale and that the problem stems from the fact that we are neglecting heavy degrees of freedom that become intertwined with the axion. 

To this end, consider quantum chromodynamics (QCD)\footnote{In this letter we use the term QCD for any QCD-like theory with strong dynamics in the IR.} with $2$ light fundamental quarks along with an axion and take the axion constant $f$ to be much larger than the strong scale. The low energy chiral Lagrangian is given by \cite{Kim:1986ax,diCortona:2015ldu}:
\begin{eqnarray}
\nonumber
{\cal L}&=&\frac{f_{\pi}^2}{4}\mbox{tr}\left[\partial_\mu U^\dagger \partial^\mu U\right]+\frac{1}{2}\left(\partial_\mu a\right)^2\\
&+&\frac{B_0 f_{\pi}^2}{2}\mbox{tr}\left[U {\cal M}^{\dagger}+{\cal M} U^{\dagger} \right]\,,
\label{chiral Lagrangian}
\end{eqnarray}
where $U=e^{i\frac{\bm \pi \cdot \bm \tau}{f_\pi}}$, $\bm \pi=(\pi^+, \pi^-, \pi^0)$,  $\bm \tau=(\tau^+,\tau^-,\tau^3)$ are the Pauli matrices,   $B_0=\frac{m_\pi^2}{(m_u+m_d)}$, and $f_\pi$, $m_\pi$, $m_u$, $m_d$ are the pion decay constant, pion mass, and up and down quark masses, respectively. $\cal M$ is the mass matrix: ${\cal M}=\mbox{diag}\left(m_u e^{i\frac{a}{f}},m_d e^{-i\frac{a}{f}} \right)$. If, for simplicity, we assume  that $m_u=m_d$,  then  we obtain the potential 
\begin{eqnarray}
\nonumber
V=\frac{B_0 f_{\pi}^2}{2}\mbox{tr}\left[U {\cal M}^{\dagger}+{\cal M} U^{\dagger} \right]=m_\pi^2 f_\pi^2 \cos\frac{\pi^0}{f_\pi}\cos\frac{a}{f}\,.\\
\label{pion axion potential}
\end{eqnarray}
If we are interested in the low-energy axion physics, then we can completely neglect the pions since $f_\pi\ll f$, which leads to the axion potential 
\begin{eqnarray}
V=m_\pi^2 f_\pi^2\cos\frac{a}{f}\,.
\label{approx axion pot}
\end{eqnarray}
This potential, apart from a cosmological constant,  is identical\footnote{We can also be sloppy and set $m_\pi \approx f_\pi\approx \Lambda$.} to (\ref{axion effective V}).   Now, let us understand what goes wrong with this potential as the field $a$ makes large excursions.  

The potential  (\ref{axion effective V})  has a unique vacuum at $a=0$. Then, one may expand the cosine near the vacuum, $V(a)\cong \Lambda^4\left(\frac{a}{f}\right)^2+...$, which is a good approximation for small perturbations  $|a|\ll f$, e.g., like when we study  scattering problems.  In other words, small perturbations near the minimum guarantees that  $|V(a)|/ \Lambda^4\ll 1$ for $|a|\ll f$ and the low-energy effective field theory (\ref{axion effective V}), which is valid for energies $E\ll \Lambda$, is robust. Now, one can immediately read the axion mass $m_a\cong \frac{\Lambda^2}{f}$, which is much lighter than the strong scale $\Lambda$. In this case we are justified to ignor the pion field, which is much heavier than the axion. The problem, however, appears once the axion makes a large  excursion $a\sim f$, as required in models of axion inflation. In this case we find $V(a)\cong \Lambda^4$ exposing the scale $\Lambda$, and the single cosine approximation (\ref{axion effective V}) is no longer trusted.  In fact, as the axion makes a large excursion, we expect that both the light (axion) and heavy (hadrons) scales to be intertwined. Hence, one needs to restore to the original potential (\ref{pion axion potential}) in order to account for the intertwining phenomenon\footnote{See \cite{Shiu:2018unx,Shiu:2018wzf} for an attempt to study  the effect of the heavy degrees of freedom on natural inflation. We also point out that in this work we consider QCD at zero temperature, in contrast with little inflation discussed in \cite{Boeckel:2009ej}, which occurs near the QCD phase transition. In \cite{Zhitnitsky:2014aja}, it was proposed that inflaton in  QCD-like models is achieved by an auxiliary nonpropagating field. This is different from the models we discuss here.}.  This complex behavior, and hence, the rearrangement of the hadronic degrees of freedom,  was also anticipated to happen, for example,  at the core of axion domain walls \cite{Fugleberg:1998kk,Halperin:1998gx,Gabadadze:2002ff,Huang:1985tt,Forbes:2000et} (where the large-field excursion takes place), based on lessons from the chiral Lagrangian, the large-$N_c$ limit, D-branes, and supersymmetry. 

It was not until recently that this scales intertwining phenomenon was put on firmer grounds by invoking a new type of 't Hooft anomaly \cite{Anber:2020xfk}. We discuss the essence of this anomaly and its implications in the next sections. In order to do that, we first embed the axion and hadrons in a UV gauge theory that entertains a plethora of global symmetries and examine the fate of the partition function as we turn on background gauge fields of these symmetries. However, we must emphasize here that whether there is an anomaly or not, the breakdown of  (\ref{axion effective V}) is expected whenever the rules of effective field theory are violated. What the anomaly buys for us is a non-perturbative statement of why the breakdown happens.

\subsection{Embedding in a UV complete gauge theory} In this section we embed the axion as well as the hadrons in a UV complete gauge theory.

 We consider an asymptotically free vector-like $SU(N_c)$ gauge theory with $N_f$ flavors of fermions in a representation ${\cal R}$ of the color group. We take the fermions to be left-handed Weyl fermions $(\psi,\tilde \psi)$ such that $\psi$ transforms under ${\cal R}$ and $\tilde \psi$ transforms under the complex conjugate representation $\bar{\cal R}$, which, in turn, guarantees the absence of gauge anomalies\footnote{The formalism requires a minor modification for fermions in self-conjugate representations, i.e.,  when $\cal R=\bar{\cal R}$, e.g., adjoint fermions. We assume throughout this work that the representation is not self-conjugate.}. We combine $\psi$ and $\tilde \psi$ into a single Dirac spinor $\Psi$. The UV Lagrangian reads
\begin{eqnarray}
\nonumber
{\cal L}_{UV}^0=\sqrt{g}\left[ \frac{1}{4g_s^2}(F_{\mu\nu}^c)^2+i\bar\Psi\slashed D \Psi \right]\,,\\
\end{eqnarray}
where the flavor and color indices are implicitly contracted, $g_s$ is the coupling constant, $g$ is the metric, and $\slashed D$ is the Dirac operator, which contains both the gauge and spin connections. 

It is essential to work out the faithful global symmetries of the theory, since they play a pivotal role in the generalized 't Hooft anomalies, as we will discuss soon.   The classical global symmetry of the theory at hand is $G^{\mbox{global}}=SU(N_f)_L\times SU(N_f)_R\times U(1)_B\times U(1)_A/\left[\mathbb Z_{N_c/p}\times\mathbb Z_{N_f}\times \mathbb Z_2\right]$. The fermions $\psi$ ($\tilde \psi$) transform under the global flavor symmetry $SU(N_f)_L\times SU(N_f)_R$ as $(\Box,1)$ ($(1,\bar\Box)$),  their charges under the baryon symmetry $U(1)_B$ is $+1$ ($-1$), while their charges under the $U(1)_A$ axial symmetry  is $+1$ ($+1$). We also  mod out by various discrete symmetries in order to avoid redundancy. These discrete symmetries can be absorbed in combinations that involve $SU(N_c)$ as well as global transformations of $U(1)_B$ and $SU(N_f)_L\times SU(N_f)_R$.  The $\mathbb Z_{N_f}$ symmetry is the center of $SU(N_f)_{L,R}$, while $\mathbb Z_2$ is the fermion number. The discrete symmetry $\mathbb Z_{N_c/p}$, on the other hand, needs more elaboration. The N-ality $n_c$ of a representation $\cal R$ is the number of the boxes in Young tableau modulo $N_c$.  Thus, the part of the gauge group that acts faithfully on fermions\footnote{One way to identify the faithful discrete symmetry $\subseteq\mathbb Z_{N_c}$  that acts on fermions  is to consider the transition functions  $\Phi_{ij}$  of the $SU(N_c)$ gauge bundle on the overlap between two patches $U_i\cap U_j$ that cover the manifold. Then, fermions of N-ality $n_c$ transform as $\psi_i\rightarrow e^{i2\pi\frac{n_c}{N_c}}\psi_j$. Hence, the fermions are blind to a $\mathbb Z_p$ transformation, where  $p=\mbox{gcd}(N_c,n_c)$.} is $SU(N_c)/\mathbb Z_p$, where $p=\mbox{gcd}(N_c,n_c)$. This means that the fermions are charged under $\mathbb Z_{N_c/p}\subseteq\mathbb Z_{N_c}$, and hence, we mod  it out since it is part of the gauge group. 

The quantum corrections break\footnote{\label{axial anomaly} One can see this breaking by studying the triangle diagrams of the $U(1)_A[SU(N_c)]^2$ anomaly. The  anomaly contributes a phase $e^{i2\alpha N_f T_{\cal R}\int \mbox{tr}_\Box\left(\frac{F^c\wedge F^c}{8\pi^2}\right)}$ to the Euclidean partition function, where $\alpha$ is the phase of the $U(1)_A$ global transformation. Thus, $U(1)_A$ is anomalous in the background of the color field, and only those values of $\alpha$ that satisfy $\alpha=\frac{2\pi k}{2N_f T_{\cal R}}$, $k \in \mathbb Z$, leave the partition function invariant.} $U(1)_A$ down to the non-anomalous discrete chiral symmetry $\mathbb Z^{d\chi}_{2N_f T_{\cal R}}$, where $T_{\cal R}$ is the Dynkin index. We normalize $T_{\cal R}$  such that the trace in the fundamental representation is $T_\Box=1$ and the simple roots $\bm \alpha$ of $SU(N_c)$ have length square $\bm \alpha^2=2$. 
Thus, the good global symmetry of our theory is reduced to 
\begin{eqnarray}
\nonumber
G^{\mbox{global}}=\frac{SU(N_f)_L\times SU(N_f)_R\times U(1)_B\times \mathbb Z^{d\chi}_{2N_f T_{\cal R}}}{\mathbb Z_{\frac{N_c}{p}}\times\mathbb Z_{N_f}\times\mathbb Z_2}\,.\\
\label{global G}
\end{eqnarray}

We further couple the fermions to a complex Higgs field $\Phi=\phi_1+i\phi_2$ by introducing  the Yukawa term $y\bar \Psi(\phi_1+i\phi_2 \gamma^5)\Psi$ and the Higgs potential $V(\Phi)=\lambda\left(|\Phi|^2-f^2\right)^2$.  The full Lagrangian reads
\begin{eqnarray}
\nonumber
{\cal L}_{UV}&=&{\cal L}_{UV}^0+\sqrt{g}\left[|\partial_\mu\Phi|^2+ V(\Phi)+ y\bar \Psi(\phi_1+i\phi_2 \gamma^5)\Psi\right]\,.\\
\label{full UV lag}
\end{eqnarray}
 The  axion constant $f$ is taken to be much larger than the strong-coupling scale $\Lambda$, i.e., we demand $f\gg \Lambda$,  and the dimensionless couplings $\lambda$ and $y$ are ${\cal O}(1)$ constants. We will also assume that $\Lambda\ll M_P$, while we comment on the relation between $f$ and $M_P$ below. The Higgs field is inert under all symmetries except $\mathbb Z^{d\chi}_{2N_f T_{\cal R}}$ and the Yukawa term is invariant under $SU(N_c)\times G^{\mbox{global}}$. The fermions and Higgs fields  transform as $\psi\rightarrow e^{-i\frac{2\pi}{2N_fT_{\cal R}}} \psi$, $\tilde\psi\rightarrow e^{-i\frac{2\pi}{2N_fT_{\cal R}}} \tilde\psi$, $\Phi\rightarrow e^{4i\frac{2\pi}{2N_fT_{\cal R}}}\Phi$ under $\mathbb Z^{d\chi}_{2N_f T_{\cal R}}$. 
 
At energy scale $\Lambda\ll E\ll f$ the Higgs field acquires a vacuum expectation value. We write $\Phi\equiv\rho e^{ia}$ and set $\rho=f$, where $a$ is the axion field. Then, the fermions acquire a mass ${\cal O}(f)$ and decouple. The effective Lagrangian becomes:
 \begin{eqnarray}
 \nonumber
 {\cal L}_{\Lambda\ll E\ll f}&=&\sqrt{g}\left( \frac{1}{4}(F_{\mu\nu}^c)^2+f^2(\partial_\mu a)^2\right)\\
 &&+a N_f T_{\cal R} \mbox{tr}_\Box\left(\frac{F^c\wedge F^c}{8\pi^2}\right) \,,
 \label{int lag}
 \end{eqnarray}
where $\mbox{tr}_\Box\left(\frac{F^c\wedge F^c}{8\pi^2}\right)$ is the topological charge density of the color field and the pre-coefficient $N_f T_{\cal R}$ is obtained by integrating out the fermions running inside the triangle diagrams that contribute to the $U(1)_A\left[SU(N_c)\right]^2$ anomaly, see Footnote \ref{axial anomaly}. The Lagrangian (\ref{int lag}) describes $N_c^2-1$ gluonic degrees of freedom coupled to an axion. Since the axion constant is much larger than the strong scale, the axion field does not experience a large variation over length scales $\sim \Lambda$, and we can think of $a$ as a constant $\theta$ angle over such length scales. Notice also that the term $a N_f T_{\cal R} \mbox{tr}_\Box\left(\frac{F^c\wedge F^c}{8\pi^2}\right)$ breaks the axion shift symmetry down to  $\mathbb Z_{N_f T_{\cal R}}$. This is an exact symmetry of the system, which remains a good one deep in the IR (but it can break spontaneously, e.g., we can have axion domain walls). 

Further, going down to energies $E\ll \Lambda$, the theory generates a mass gap and confines\footnote{Strictly speaking, true confinement will only take place if the system has an unbroken $\mathbb Z_{N_c}^{1}$ 1-form center symmetry. This symmetry acts on the Polyakov loop and guarantees that the vacuum expectation value of the  loop is zero in the confining regime, i.e., we have an infinitely long flux tube between two fundamental probe charges. A theory with dynamical fundamental fermions, for example, does not have a  $\mathbb Z_{N_c}^{1}$  symmetry, and thus, there is no true notion of confinement. However, since $f\gg \Lambda$, the flux tubes break at a length scale $\sim \frac{f}{\Lambda^2}$, which is parametrically much larger than $\Lambda^{-1}$. In this case, we can talk about an {\em emergent} $\mathbb Z_{N_c}^{1}$ 1-form symmetry.}. The only relevant degree of freedom, then, is the axion field. Integrating out the strong field fluctuations generates a potential\footnote{The potential is generated after summing over a dilute gas of  Belavin-Polyakov-Schwarz-Tyupkin (BPST) instantons. Every (anti)instanton carries a minimum topological charge of $\int \mbox{tr}_\Box\left(\frac{F^c\wedge F^c}{8\pi^2}\right)=\pm1$. Thus, the 't Hooft vertex in the background of the BPST instanton takes the form $e^{-\frac{8\pi^2}{g_s^2}}e^{\pm i a N_f T_{\cal R}}$, where $\frac{8\pi^2}{g_s^2}$ is the instanton action. The 't Hooft vertex is invariant under the discrete shift symmetry $a\rightarrow a+\frac{2\pi}{N_f T_{\cal R}}$. Inserting the vertex into the path integral generates the potential $V(a)$. However, one has to keep in mind that the dilute gas approximation is not under control since the scale modulus grows indefinitely, thanks to the strong-coupling behavior in the IR.} $V(a)$:
\begin{eqnarray}
\nonumber
{\cal L}_{E\ll \Lambda}&=&\sqrt{g}\left[\frac{1}{2} \left(\partial_\mu a\right)^2+V(a) \right]\,,\\
V(a)&=&\Lambda^4 \left(1-\cos\left(\frac{a N_f T_{\cal R}}{f}\right)\right)+...\,,
\label{axion Lagrangian}
\end{eqnarray}
where we scaled $a\rightarrow a/f$. The dots in (\ref{axion Lagrangian}) refer to higher harmonics\footnote{The higher harmonics can be thought of  summing over BPST instantons with higher topological charges.}, which need to respect the discrete shift symmetry $a\rightarrow a+\frac{2\pi}{N_f T_{\cal R}}$. As we pointed out above, the potential (\ref{axion Lagrangian}) cannot be trusted when the axion makes large-field excursions; the effective field theory breaks down. We will also show that this potential is inconsistent with a new 't Hooft anomaly that we will discuss momentarily. Before doing that, we digress to discuss one technical aspect that has to do with the value of the axion constant.

\subsection{N-flation} The potential in (\ref{axion Lagrangian}) is typical in studying models of axion inflation. In order to satisfy the Planck satellite constraints on the CMB power spectrum \cite{Akrami:2018odb}, however, one needs to take $f>M_P$, which is in tension with theories of quantum gravity\footnote{For example, if one takes $f>M_P$, then gravitational instantons can induce higher harmonics that spoil inflation, see, e.g., \cite{Montero:2015ofa}.}. The tension can be alleviated, for example, by  invoking the idea of N-flation \cite{Dimopoulos:2005ac}, where we postulate the existence of $N$ axion fields\footnote{Another mechanism that can be used to circumvent the tension is the aligned natural inflation \cite{Kim:2004rp,Peloso:2015dsa}.}. Each field has an axion constant $f<M_P$. Interestingly, one finds that the collective motion of the axions results in an effective axion constant $f_{\scriptsize\mbox{eff}}\cong \sqrt{N}f>M_P$, and hence, we can respect the CMB constraints without going against the general lore we learn from theories of quantum gravity. In order to illustrate this mechanism, we briefly repeat the above steps, but now we couple the fermions to $N$ complex Higgs fields: $\sum_{j=1}^N \bar\Psi\left(\phi_1^j+i\phi_2^j\gamma^5\right)\Psi$, and introduce the Higgs potential $V=\sum_{j=1}\lambda\left(|\Phi_j|^2-f^2\right)$. We immediately find  that each axion $a_j$ respects the discrete shift symmetry $a_j\rightarrow a_j+\frac{2\pi}{N_f T_{\cal R}}$. Integrating out the fermions generates the term $\sum_{j=1}^Na_j \mbox{tr}_\Box\left(\frac{F^c\wedge F^c}{8\pi^2}\right)$, while summing over the instantons\footnote{There will be $N$ distinct 't Hooft vertices $\sim e^{-\frac{8\pi^2}{g_s^2}}e^{\pm i a_j N_f T_{\cal R}}$ for $j=1,2,...,N$. Notice that 't Hooft vertices are invariant under the discrete shift symmetry $a_j\rightarrow a_j+\frac{2\pi}{N_f T_{\cal R}}$.} yields the low-energy  effective Lagrangian  
\begin{eqnarray}
\nonumber
{\cal L}_{E\ll \Lambda}&=&\sqrt{g}\left[\frac{1}{2} \sum_{j=1}^N\left(\partial_\mu a_j\right)^2\right.\\
&&\left.+\sum_{j=1}^N\Lambda^4  \left(1-\cos\left(\frac{a_j N_f T_{\cal R}}{f}\right)\right) \right]\,.
\end{eqnarray}
In order to simplify the analysis, we assume that the $N$ distinct axions have the same initial conditions as they start to roll down the potential, i.e., $a_j\cong \pi f/(N_f T_{\cal R})$. Thus, we take $a_j=a$ for every $j=1,2,...,N$ and define $a_e\equiv \sqrt{N}a$ to obtain
 \begin{eqnarray}
 \nonumber
{\cal L}_{E\ll \Lambda}^{\scriptsize\mbox{eff}}&=&\sqrt{g}\left[\frac{1}{2} \left(\partial_\mu a_e\right)^2\right.\\
&&\left.+N\Lambda^4 \left(1-\cos\left(\frac{a_e N_f T_{\cal R}}{\sqrt{N}f}\right)\right)\right]\,,
\end{eqnarray}
and the collective degree of freedom $a_e$ has an effective axion constant $\sqrt{N}f>M_P$ for large enough values of $N$. What we have achieved is that each axion has a large-field excursion less than the Planck scale, yet, the effective axion $a_e$ yields scalar and tensor perturbations that are compatible with the CMB power spectrum \footnote{It can also be shown that the effect of gravitational instantons is suppressed in models of N-flation and that these models are not in conflict with the weak gravity conjecture, see \cite{Montero:2015ofa}.}.

From now on, we work with the Lagrangian (\ref{axion Lagrangian}), keeping in mind that the axion field and the axion constant that appear there are the effective ones. Now, we return back to the original problem and address the fact that the Lagrangian (\ref{axion Lagrangian}) breaks down as $a$ makes large excursions. In order to remedy the problem, we need to introduce the strongly-coupled degrees of freedom  into the effective potential, as we discussed above in the case of the chiral Lagrangian. Here, unfortunately, one cannot use the power of chiral perturbation theory since the fermions are heavy and decouple long before the theory enters its strongly coupled regime. In the next sections we discuss how to address this problem and introduce the reader to a recently discovered 't Hooft anomaly that enables us to make non-trivial statements about the IR physics. 

\subsection{Generalized 't Hooft anomalies: the BCF anomaly} An 't Hooft anomaly is an obstruction to gauging a global symmetry \cite{tHooft:1979rat}. The anomaly is a renormalization group invariant, and hence, it has to be matched between the UV and IR\footnote{As a word of caution, the reader should not confuse this kind of anomalies with gauge or axial anomalies. If a theory has a gauge anomaly, then it is either sick in the UV or can make sense only as an effective field theory with a cutoff, see \cite{Preskill:1990fr}. An example of the axial anomaly is the one we discussed in Section \ref{Theory and formulation}, where $U(1)_A$ is broken down to a discrete subgroup in the background of the dynamical color field. 't Hooft anomaly, on the other hand, neither signals anything wrong with the theory, nor does it reduce the global symmetry. It is just a way to probe the theory by gauging its global symmetries.}. This makes 't Hooft anomalies an indispensable tool to study asymptotically free theories since one can calculate the anomaly coefficient in the UV and, irrespective of the details of  dynamics, this coefficient has to be exactly reproduced by the IR strongly-coupled degrees of freedom.  These anomalies were known since the 80's and played an important role in model building of composites \cite{Rosner:1998wh} and Seiberg dualities \cite{Seiberg:1994pq}. Recently, this topic gained momentum due to the discovery of new anomalies that go under the name {\em generalized 't Hooft anomalies}, see, e.g., \cite{Gaiotto:2014kfa,Gaiotto:2017yup,Anber:2019nze,Anber:2020gig} and references therein.

Due to the limited space, we only give a glimpse of the essence of these new anomalies. The interested reader is referred to the cited literature for more details. 

 The traditional 't Hooft anomalies, as were know since the 80's, involve only {\em integer} topological charges of the background gauge field of a given global symmetry.  The new realization,  that was first made in \cite{Gaiotto:2014kfa}, is that one can also turn on {\em fractional} topological charges known as 't Hooft fluxes \cite{tHooft:1979rtg}. This new technology was applied in \cite{Gaiotto:2017yup} to gauging  the center symmetry of the color group. Recently,  this method was generalized by turning on fractional  fluxes in the color, flavor, and baryon number directions, compatible with the faithful action of the global symmetry of a given theory \cite{Anber:2019nze,Anber:2020gig}. The new anomaly was dubbed the {\em BCF anomaly}.  

In fact, by closely examining the microscopic theory discussed in Section \ref{Theory and formulation}, one reveals that it exhibits a BCF anomaly. This is a mixed anomaly between the global discrete chiral symmetry $\mathbb Z^{d\chi}_{2N_f T_{\cal R}}$ and the color-baryon-flavor background fluxes. These are the most general fluxes that are compatible with the faithful global symmetry of the theory as given in (\ref{global G}).   Again, we spare the reader the details and refer to \cite{Anber:2019nze} for a comprehensive explanation of all the steps. Succinctly, we compactify the Euclidean theory on a large $4$-torus\footnote{The compactification on a $4$-torus is a slick way to see the anomaly. However, we must emphasize that the anomaly itself is insensitive to the details of the geometry.}. Then,  the BCF anomaly shows up in the UV as an irremovable phase\footnote{This means that we cannot introduce a counter term that removes the phase.} in the Euclidean partition function ${\cal Z}$ as we perform a $\mathbb Z^{d\chi}_{2N_f T_{\cal R}}$ global transformation in the  color-baryon-flavor background fluxes:
\begin{eqnarray}
{\cal Z}\longrightarrow  {\cal Z} e^{i \frac{2\pi}{N_f T_{\cal R}}\left[N_f T_{\cal R}Q^c+ d_{\cal R}\left(Q^f+Q^B\right) \right]}\,,
\label{UV anomaly}
\end{eqnarray}
where $d_{\cal R}$ is the dimension of the representation ${\cal R}$ and $Q^{c,B,f}$ are the fractional topological charges in the color, baryon, and flavor directions. These charges are given by:
\begin{eqnarray}
\nonumber
Q^c&=&mm'\left(1-\frac{1}{N_c}\right)\,, Q^f=kk'\left(1-\frac{1}{N_f}\right)\,,\\
Q^B&=&\left(n_c\frac{m}{N_c}+\frac{k}{N_f}\right)\left(n_c\frac{m'}{N_c}+\frac{k'}{N_f}\right)\,,
\label{fractional Q}
\end{eqnarray}
where $m,m',k,k'\in\mathbb Z$. At energy scale $\Lambda\ll E\ll f$, the BCF anomaly can be reproduced by coupling the axion to the topological charge densities of the color, flavor, and baryon number. Thus, we need to add the following term 
\begin{eqnarray}
L_{\scriptsize\mbox{anom}}=a\left(N_fT_{\cal R}q^c+d_{\cal R}q^f+d_{\cal R}q^b\right)\,,
\label{anomaly IR lag}
\end{eqnarray}
to the infrared Lagrangian in order to match the BCF anomaly. Here, $q^{c,B,f}$ are the topological charge densities of the color, flavor, and baryon number, $q^{c,B,f}=\mbox{tr}_\Box\left(\frac{F^{c,B,f}\wedge F^{c,B,f}}{8\pi^2}\right)$, whose integrals on a closed 4-D manifold give $Q^{c,B,f}$ in (\ref{fractional Q}). Now, we see that under a discrete shift symmetry $a\rightarrow a+ \frac{2\pi}{N_f T_{\cal R}}$ the IR partition function acquires the exact same UV phase that appears in Eq. (\ref{UV anomaly}). 

The Lagrangian (\ref{anomaly IR lag}) does not immediately dictate the mathematical form of the deep IR effective field theory at energy scale $\ll \Lambda$. The dynamics in the IR, however, has to conspire in order to respect the BCF anomaly: this anomaly involves the color direction, and hence, the IR effective field theory has to contain information about the color degrees of freedom\footnote{The BCF anomaly is unlike other ordinary (associated to integer fluxes) 't Hooft anomalies. For example, the theory we are discussing enjoys two ordinary 't Hooft anomalies: $\mathbb Z_{2N_f T_{\cal R}}^{d\chi}\left[U(1)_B \right]^2$ and $\mathbb Z_{2N_f T_{\cal R}}^{d\chi}\left[SU(N_f) \right]^2$. These anomalies, however, do not involve the color field, and thus, they do not tell us about any kind of intertwining between the axion and strongly coupled theory in the IR.}.  Thus, the single-field Lagrangian (\ref{axion Lagrangian}), that is used in models of axion inflation, is missing important information about this anomaly and it cannot be the whole story.   The very interesting point, though, is that this information is related to a very high energy scale $\sim\Lambda$ compared to the scale of axion physics $\sim \Lambda^2/f$; this information needs to lurk deep in the IR in order to match the BCF anomaly. This is the nonperturbative reason of the scale-intertwining phenomenon that was anticipated long ago  \cite{Fugleberg:1998kk,Halperin:1998gx,Gabadadze:2002ff}. However, we must be clear that whether the anomaly is present or not, one still expects the strong dynamics to play a major role as the axion makes a large-field excursion\footnote{\label{note fund}In fact, a theory with a single  fundamental quark does not have a genuine discrete chiral symmetry. Hence, in this case there is no BCF anomaly. Yet, one expects the strong dynamics to play a role as the axion makes a large-field excursion.}. As we argued above, the effective potential (\ref{axion Lagrangian}) breaks down in this case since the hadronic degrees of freedom get excited. What the anomaly buys for us, though, is a nonperturbative statement why the interplay between hadrons and axion should take place. 

The question then is, how do we account for the strong dynamics in the IR? Given the poor handle we have on the strongly-coupled phenomena, this sounds like a daunting task and there is not much one can say about the effects of the strong dynamics on axion inflation. Nevertheless, over the past decade we have learned a good deal of technology that enables us to understand the physics of the strong dynamics, e.g., confinement and discrete chiral symmetry breaking, by means of weakly coupled physics. We review the progress in this direction in the next section and set the stage to investigate axion inflation in a more realistic setup compared to what has been achieved in the literature. 

\subsection{Deformed QCD} In order to have  control over the IR dynamics, we deform the original theory given by the Lagrangian (\ref{full UV lag}) such that we leave the global symmetry (\ref{global G}) intact. This method of deforming a strongly-coupled gauge theory has been invoked for more than a decade with a huge body of literature. We refer the reader to the review \cite{Dunne:2016nmc}, while we try to keep our discussion succinct and relevant to the cosmological context. At the end of this section, we remove the  deformation and motivate a model that takes into account the effects of the strong dynamics on axion inflation. 

To this end, we compactify the theory\footnote{Remember that the IR theory is in a confining regime.}  on a spatial circle $\mathbb S^1_L$ of circumference $L$ and at the beginning we take $L$ to be much larger than any length scale in the problem\footnote{This is not a thermal circle, and hence, deformed QCD is not a finite temperature theory.} and  add the {\em double-trace deformation}
\begin{eqnarray}
L_{\scriptsize{\mbox{DTD}}}=\sum_{n=1}c_{n}\mbox{Tr}_\Box|\Omega^n|^2\,.
\label{DTD}
\end{eqnarray}
We say that the theory lives on $\mathbb R^{2,1}\times \mathbb S^1_L$, where $\mathbb R^{2,1}$ denotes two space and one time directions\footnote{Here, we slightly deviate from the promise that was made throughout this section that we are woking in a Euclidean version of the theory in order to emphasize that $\mathbb S_L^1$ is not a thermal circle.}. However, given that the length of the circle is very large, the dynamics does not distinguish between this theory and the theory that lives on $\mathbb R^4$; it   is still in the confining regime in the IR.
$\Omega\equiv e^{i\oint_{\mathbb S^1_L}A^c}$ is the Polyakov loop (or holonomy) of the color field over $\mathbb S^1_L$  (the color field obeys periodic boundary conditions over $\mathbb S^1_L$) and $c_n$ are positive coefficients  $\sim {\cal O}(1)$, as we explain below.  We also give the fermions and Higgs field periodic  boundary conditions\footnote{Actually, whether we give them periodic or anti-periodic boundary conditions will not affect our analysis since they decouple at scales much higher than the strong scale.} along $\mathbb S^1_L$ .

Next, we  lower $L$ below $1/\Lambda N_c$; we take $\Lambda\ll\frac{1}{N_c L} \ll f$. In this case\footnote{Since $\frac{1}{L}\ll f$, the fermions and radial component of the Higgs are already decouple.} the gauge field fluctuations generate a potential for the holonomy $\Omega$. This potential favors a center-broken phase, $\mbox{Tr}_\Box\Omega\neq 0$; this is the celebrated thermal phase transition (deconfinement) that happens at $L\sim \Lambda^{-1}$. In order to suppress the gauge field fluctuations, avoid the transition, stay in a center-symmetric confining phase ($\mbox{Tr}_\Box\Omega=0$),  and allow for a smooth behavior across $L\sim \Lambda^{-1}$, we choose the coefficients $c_n$ in (\ref{DTD}) to be large and positive. In summary, adding the double-trace deformation to the theory allows for an adiabatic continuity between large and small values of $L$ and guarantees that the theory stays in its confining regime. Next, we study the dynamics of the theory in the small circle limit, which is totally under analytical control, and show how confinement happens. This task is not possible on $\mathbb R^4$ or in the limit $\Lambda\gg\frac{1}{N_c L}$.

Definitely, a small value of $L$ is not realistic for the cosmological context. What we are trying to do here is motivate a model that can capture the physics of the strong dynamics  as we take $L\rightarrow \infty$. We argue below that such a model is consistent with what is expected in a confining theory.

Let us for now continue our investigation of the deformed theory on the circle.  If we take $L$ to be small enough, $L\Lambda N_c\ll 1$, then the theory enters its weakly coupling regime, $g_s\ll1$, and becomes amenable to the semi-classical techniques\footnote{Upon the breakdown of $SU(N_c)$ down to $U(1)^{N_c-1}$, see the few lines that follow, we obtain a tower of Kaluza-Klein  W-bosons. The lightest  W-boson mass is $\frac{2\pi}{N_cL}$. Thus, when we take  $\Lambda N_c\ll 1$ the W-boson mass works as an IR cutoff that stops the running of the coupling constant $g_s$ at energy $\sim\frac{1}{N_cL}$, which is much higher than the strong scale, and hence, the theory is weakly coupled at this scale.}.  Because the theory is at a center-symmetric point, the gauge field component in the compact direction, which behaves as a scalar, gets a non-zero vacuum expectation value. Then,  the gauge group $SU(N_c)$ spontaneously breaks  into the maximal abelian subgroup $U(1)^{N_c-1}$. Effectively, the theory lives in $3$-D and each of the  $U(1)^{N_c-1}$ photons can be dualized to a scalar. The photon fields can be taken along the directions of the Cartan subalgebra. Let the bold-face symbol $\bf F$ denote the photon fields along the Cartan subalgebra directions: ${\bm F}=(F_1,F_2,...,F_{N_c-1})$. Then, the duality relation in $3$-D reads ${\bm F}_{\mu\nu}\sim\epsilon_{\mu\nu\alpha}\partial_\alpha\bm \sigma$, where $\bm \sigma$ is the dual photon. At this stage, the reader might conclude that the theory has $N_c-1$ massless degrees of freedom, and thus, it is dramatically different from the mother theory before compactification\footnote{Remember that we are supposed to be in a confining regime (this is the essence of the adiabatic continuity) and the number of effective degrees of freedom should scale as $(N_c)^0$.}. This conclusion, however, is premature since the theory also admits monopole-instantons\footnote{The monopole-instantons are the constituents of a BPST instantons, and they are reliable saddle points of the path integral in the semi-classical limit.}. The 't Hooft vertices of these instantons and anti-instantons are
\begin{eqnarray}
\nonumber
{\cal M}_k&=&e^{-\frac{8\pi^2}{Ng_s^2}}e^{i\bm \alpha_k\cdot \bm \sigma+i\frac{N_f T_{\cal R}}{N_c}a}\,,\\
 \bar{\cal M}_k&=&e^{-\frac{8\pi^2}{Ng_s^2}}e^{-i\bm \alpha_k\cdot \bm \sigma-i\frac{N_f T_{\cal R}}{N_c}a}\,,
\end{eqnarray}  
for  the simple and affine roots $\bm \alpha_k$, $k=1,2,...,N_c$.
In the weak-coupling regime the monopoles are a dilute gas, and their effect can be taken into account by summing over an ensemble of them in the partition function. The final expression of the $3$-D effective Lagrangian reads: 
\begin{eqnarray}
\nonumber
L_{\scriptsize\mbox{eff}}^{3-D}=\sqrt{g}\left[\frac{g_s^2}{8\pi^2 L}\left(\partial_\mu\bm \sigma\right)+\frac{f^2 L}{2}\left(\partial_\mu a\right)^2+V(\bm \sigma,a)\right]\,,\\
\label{Lagrangian on a small circle 1}
\end{eqnarray}
where
\begin{eqnarray}
\nonumber
V(\bm \sigma,a)=\frac{1}{L^3}e^{-\frac{8\pi^2}{N_c g^2}}\sum_{k=1}^{N_c}\left(1-\cos\left(\bm \alpha_k\cdot \bm \sigma+\frac{ N_f T_{\cal R}}{N_c}a\right)\right)\,.\\
\label{Lagrangian on a small circle}
\end{eqnarray}
Therefore, we see that the monopole-instantons generate a mass gap for the photons, as can be easily seen by expanding the cosine term to quadratic order, which leads to the confinement of the fundamental quarks.  It is crucial to note that the potential (\ref{Lagrangian on a small circle}) is valid for both small and large-field excursions of $a$ and $\bm \sigma$, and its validity is guaranteed as long as we are in the semi-classical regime $L N_c\Lambda\ll1$.

It can also be shown that the BCF anomaly of the UV theory carries over in the deformed QCD version of the theory, as was explained in details in  \cite{Anber:2019nze} and we do not repeat this discussion here. It suffices to say that the fractional nature of the background topological charges (\ref{fractional Q}) plays a pivotal role when compactifying the theory on a small manifold. A more technical way of thinking about the fractional charges is to turn on higher-form background fields and couple our theory to a topological quantum field theory \cite{Kapustin:2014gua}. Upon compactifying a theory on a circle, such background fields persist in the small circle limit, leading to the same BCF anomaly of the original theory  \cite{Anber:2019nze}. Interestingly, integral topological charges do not necessary survive the small circle compactification\footnote{To be more precise, whether the anomalies that correspond to integer topological charges survive the compactification or not is an open question.}, and hence, the fractional charges are crucial to preserve the information about the UV theory. 

The Lagrangian (\ref{Lagrangian on a small circle}) describes the interesting intertwining phenomenon between the axion and the massive photons. It is easy to see that the ratio between the photon and axion masses is $\frac{m_\sigma}{m_{a}}\sim fL\gg 1$, and one might be tempted to integrate out the photon field. This, however, overlooks the importance of the photon field in the case of a large axion excursion. For example, it was shown in \cite{Anber:2019nze} that the fundamental quarks are deconfined on axion domain-walls, thanks to the intertwining between the photon and axion fields at the core of the walls. This phenomenon is also expected to persist in the large circle limit, since the BCF anomaly is insensitive to the size of the manifold. However, the naive potential (\ref{axion Lagrangian}) cannot account for it because of the absence of any strong-dynamics information. Now, in the light of the BCF anomaly, we see what goes wrong with  (\ref{axion Lagrangian}): this potential does not match the BCF anomaly, and hence, cannot be the whole story in the IR. 

\subsection{The decompactification limit and cosmological model} 

Strictly speaking, the Lagrangian (\ref{Lagrangian on a small circle 1}) is trusted only in the semi-classical regime $N_c\Lambda L\ll1$. Upon decompactification, higher-order Kaluza-Klein monopole-instantons become important, and therefore, one loses control over the semi-classical analysis. Nonetheless, the double-trace deformation suppresses  any potential phase transition as we vary the circle size. Over the past decade, there has been a large body of evidence that the theories on the small and large circles are continuously connected, see \cite{Dunne:2016nmc} for a review. One of the latest tests is the lattice study  in \cite{Bonati:2019unv}, where it was shown that the topological susceptibility of  pure Yang-Mills theory with double-trace deformation is independent of the circle size.  

At this stage, and in the absence of a more realistic way to take into account the effects of the strong dynamics on the axion field, we postulate the following phenomenological model\footnote{One interesting aspect of this model is the appearance of $N_c$ multiplying the axion constant $f$. Therefore, for a large number of colors one can have an effective axion constant $f_e=N_c f\gg M_P$. This observation seconds  with the fact that the model (\ref{model of axion inflation 1}) has the correct  functional dependence of the vacuum energy on the $\theta$ angle. Here, we do not use this observation to enhance the value of $f$, as was done in \cite{Yonekura:2014oja,Dine:2014hwa}. Instead, we invoke N-flation as we did above.} in the $L\rightarrow \infty$ limit, which is based on (\ref{Lagrangian on a small circle 1}) and (\ref{Lagrangian on a small circle}):
\begin{eqnarray}
\nonumber
{\cal L}_{\scriptsize\mbox{model}}&=&\sqrt{g}\left[\frac{1}{2}\left(\partial_\mu a\right)^2+\frac{1}{2}\left(\partial_\mu \bm \sigma\right)^2+V(a,\bm\sigma)\right]\,,\\
\nonumber
V(a,\bm\sigma)&=&\Lambda^4\sum_{k=1}^{N_c}\left(1-\cos\left(\frac{\bm \alpha_k\cdot \bm \sigma}{\Lambda}+\frac{ N_f T_{\cal R}}{N_c f}a\right)\right)\,.\\
\label{model of axion inflation 1}
\end{eqnarray}
If we neglect the dynamics of the axion and treat it as a constant vacuum angle, $a\rightarrow \theta$, then the vacuum energy density is given by
\begin{eqnarray}
V_0=-2N_c\Lambda^4 \mbox{max}_k\left(\cos \left(\frac{2\pi k +N_f T_{\cal R}\theta}{N_c}\right)\right)\,,
\label{vacuum energy}
\end{eqnarray}
where the max function selects the branch $k$ that minimizes the cosine. This functional dependence of the vacuum energy on $\theta$ exactly resembles what one would expect based on arguments from supersymmetric gluodynamics, D-branes, and the large-$N_c$ limit \cite{Gabadadze:2002ff}. 

The fact that the phenomenological Lagrangian (\ref{model of axion inflation 1}) gives the correct functional dependence on $\theta$ makes it a
viable playground that we can use  in order to study the effects of the hadronic physics on the inflaton and cosmological perturbations in models of natural inflation. This is exactly our task in the next section. Since we will mostly consider $SU(N_c=2)$, we give the explicit form of the potential in this case. The potential (\ref{model of axion inflation 1}) reduces to\footnote{The roots of $SU(2)$ are $\alpha=\pm \sqrt 2$. We also make the substitution $\Lambda\rightarrow \sqrt 2 \Lambda$, which introduces ${\cal O}(1)$ non essential number in front of the potential that we neglect.}
\begin{eqnarray}
V(a,\sigma)=2\Lambda^4 \left[1-\cos\frac{\sigma}{\Lambda}\cos \frac{N_f T_{\cal _R}}{2f}a \right]\,.
\label{su2 potential}
\end{eqnarray}
Interestingly, this potential is identical in form to Eq. (\ref{pion axion potential}) that was based on the chiral Lagrangian. Yet, we do not expect the two Lagrangians to be related since they are based on different physics. While the Lagrangian  (\ref{pion axion potential}) is the low energy description of a strongly-coupled theory with light quarks, Eq. (\ref{su2 potential}) is descendant from a theory with heavy fermions and compactified on a circle.  Moreover, the $\sigma$ field in (\ref{su2 potential}) models a strongly-coupled degree of freedom that is different in nature from the pion field $\bm \pi$. Add to that, in an $SU(N_c)$ theory our phenomenological model describes $N_c-1$ degrees of freedom in the IR, which cannot be justified in a true strongly-coupled theory. Nonetheless, in the next section we show that the number of colors doesn't play a major role in axion inflation, at least for small enough values of $N_c$, as we conclude from comparing the dynamics of $SU(2)$ and $SU(3)$.

\section{Dynamics of axion inflation}
\label{Dynamics of axion inflation}

In this section we study inflation in the Friedmann-Robertson-Walker spacetime $ds^2=-dt^2+b^2(t)d\bm x^2$, where $b(t)$ is the scale factor and $t$ is the cosmic time. We also study the curvature and tensor perturbations and assume that the inflaton is solely responsible for the generation of the curvature perturbations, i.e., we do not invoke curvatons.

\subsection{Axion inflation: the traditional path} 

Before studying the model (\ref{model of axion inflation 1}), we pause here in order to review the dynamics of axion inflation that is based on the traditional potential (\ref{axion Lagrangian}). Inflation happens as the axion starts anywhere near the hilltop $\frac{a}{f}\cong \pi$ and rolls down to the bottom $a\cong 0$ (we can always restrict the motion of $\frac{a}{f}$ in the interval $[0,\pi]$ without loss of generality). Then,  it is a straightforward exercise to calculate the number of e-folds: $N_e=\int_{t_i}^{t_f} dt H$, where $H$ is the Hubble parameter, $t$ is the cosmic time, and the integration spans the time period from the beginning, $t_i$, to end, $t_f$, of inflation\footnote{Remember that we need $N_e\sim 50-60$ in oder to solve the problems of the standard Big-Bang cosmology.}. Then writing $\int_{t_i}^{t_f} dt H$ as $\int_{a_i}^{a_f} da  \frac{H}{\dot a}$, using the approximate equation of motion $3H \dot a+\frac{\Lambda^4}{f}\sin\frac{a}{f}=0$, where we have neglected the second derivative $\ddot a$ (assuming that the axion slowly rolls down its potential), and Friedmann's equation $3M_P^2H^2\cong V(a)$, we find $N_e=\frac{2 f^2}{M_P^2}\log \frac{\cos \frac{a_f}{2f}}{\cos \frac{a_i}{2f}}$. We also use the same approximation to calculate the slow-roll parameters $\epsilon\equiv -\frac{\dot H}{H^2}$ and $\eta\equiv\frac{\partial ^2 V/\partial a^2}{3H^2}$ to find: $\epsilon=\frac{M_P^2}{2 f^2}\mbox{cotan}^2\frac{a}{2f}$ and $\eta=\frac{M_P^2}{ f^2}\frac{\cos \frac{a}{f}}{1-\cos\frac{a}{f}}$. Inflation ends when either $\epsilon\sim 1$ or $|\eta|\sim1$. It is easy to have many e-folds of inflation if we take\footnote{Remember that theories of quantum gravity are in tension with taking $f>M_P$. Here, we assume that we are working within the N-flation model, as we discussed before.} $f>M_P$.  Interestingly,  even for values of $f< M_P$, one can also achieve a very large number of e-folds by starting  inflation very close\footnote{Strictly speaking, we cannot start inflating arbitrary close to $a=\pi f$ in order to avoid the quantum kicks, which change the vacuum expectation value of $a$ by $H$ every Hubble time. Instead, we should start at least a distance $H$ from $a=\pi f$. We would like to thank Lorenzo Sorbo for emphasizing this point.} to $\pi$. Nevertheless, recent constraints from Planck  satellite put severe constraints on the spectral tilt \footnote{We remind the reader that the power spectrum of the curvature perturbations is given by ${\cal P}(k)=A_s\left(\frac{k}{k^\star}\right)^{n_s-1}$, where $A_s=2.196\times 	10^{-9}$ and the pivot scale $k^\star=0.05~\mbox{Mpc}^{-1}$ for Planck satellite.} $n_s-1$ of the power spectrum and the scalar to tensor ratio $r$. Bounds on $n_s$ from Planck plus WMAP are $0.9457<n_s< 0.9749$ at 95\% CL \cite{Akrami:2018odb,Pajer:2013fsa}. One finds that we need $f> 10 M_P$ in order to respect this constraint.  Let us also mention that the constraints on the amplitude of the CMB power spectrum can be met by taking $\Lambda\sim 10^{-3}M_P$.

Taking $f>M_P$, one can safely assume that inflation ends at $a\sim 0$. In this case the analysis simplifies and we express both $\epsilon$ and $\eta$ as functions of the e-folds remaining before the end of inflation $N_e^\star$\footnote{In order to derive Eq. (23) we use $N_e=\frac{2f^2}{M_P^2}\log \frac{\cos \frac{a_f}{2f}}{\cos \frac{a_i}{2f}}$ and set $a_f=0$ when $N_e=0$; thus the number of remaining e-folds  just before the end of inflation is $0$. From this we find $N_e^\star=-\frac{16\pi^2 f^2}{M_P^2}\log\cos \frac{a_i}{2f}$, which gives the number of the remaining e-folds before the end of inflation, at the time when $a$ is equal to $a_i$. Then, solving for $a_i$ as a function of $N_e^\star$ and substituting into $\epsilon=\frac{M_p^2}{2f^2}\cot^2 \frac{a_i}{2f}$ we arrive to Eq. (\ref{epsilon nstar}).}: 
\begin{eqnarray}
\nonumber
\epsilon(N_e^\star)&=&\frac{M_P^2}{2 f^2}\frac{1}{e^{\frac{M_P^2 N^\star_e}{f^2}}-1}\,,\\
\eta(N_e^\star)&=&\frac{M_P^2}{2 f^2}\frac{2-e^{\frac{M_P^2 N^\star_e}{ f^2}}}{e^{\frac{M_P^2 N^\star_e}{f^2}}-1}\,.
\label{epsilon nstar}
\end{eqnarray}
Recalling that the scalar perturbations and tensor to scalar ratios are $n_s=1-6\epsilon+2\eta$ and $r=16\epsilon$, respectively, one writes both $n_s$ and $r$ as functions of $N_e^\star$:
\begin{eqnarray}
\nonumber
n_s-1&=&-\frac{M_P^2}{ f^2}\frac{e^{\frac{M_P^2 N^\star_e}{f^2}}+1}{e^{\frac{M_P^2 N^\star_e}{f^2}}-1}\,,\\
r&=&\frac{8M_P^2}{ f^2}\frac{1}{e^{\frac{M_P^2 N^\star_e}{f^2}}-1}\,.
\label{tilt and r}
\end{eqnarray}
These are the values of the spectral tilt and tensor to scalar ratio at the time when a pivot scale, probed by the CMB, excited the horizon $N_e^\star$ e-folds before the end of inflation.

\subsection{Axion inflation: the QCD effects} In this section we study the dynamics of inflation using the model (\ref{model of axion inflation 1}). We limit our investigation to the two cases $SU(N_c=2)$ and $SU(N_c=3)$. We shall find that the details of the gauge group have a little effect on the qualitative behavior of inflation. We simplify the analysis\footnote{The existence of a $\mathbb Z_{2N_f T_{\cal R}}^{d\chi}$ discrete chiral symmetry can lead to cosmological problems at the end of inflation \cite{Zeldovich:1974uw}. We ignore this problem in our analysis.} by using the replacement $f\rightarrow 2f/(N_f T_{\cal R})$ in Eq. (\ref{su2 potential}):
\begin{eqnarray}
\nonumber
{\cal L}^{SU(2)}&=&\sqrt{-g} \left[\frac{1}{2} (\partial_\mu a)^2+ \frac{1}{2} (\partial_\mu \sigma)^2+V(a,\sigma)\right]\,,\\
V(a,\sigma)&=&2\Lambda^4\left[1-\cos\frac{\sigma}{\Lambda}\cos \frac{a}{f} \right]\,.
\label{inflation su2}
\end{eqnarray}
The equations of motion and Friedmann's equation read:
\begin{eqnarray}
\nonumber
&&\ddot a+3H\dot a+\frac{\partial V(a,\sigma)}{\partial a}=0\,,\quad
\nonumber
\ddot \sigma+3H\dot \sigma+\frac{\partial V(a,\sigma)}{\partial \sigma}=0\,,\\
&&3M_P^2H^2=\frac{1}{2}\dot a+\frac{1}{2}\dot \sigma +V(a,\sigma) \,,
\label{full system}
\end{eqnarray}
and the dot denotes the derivative with respect to the cosmic time $t$.

\begin{figure}[t]
\begin{center}
\includegraphics[width=87mm]{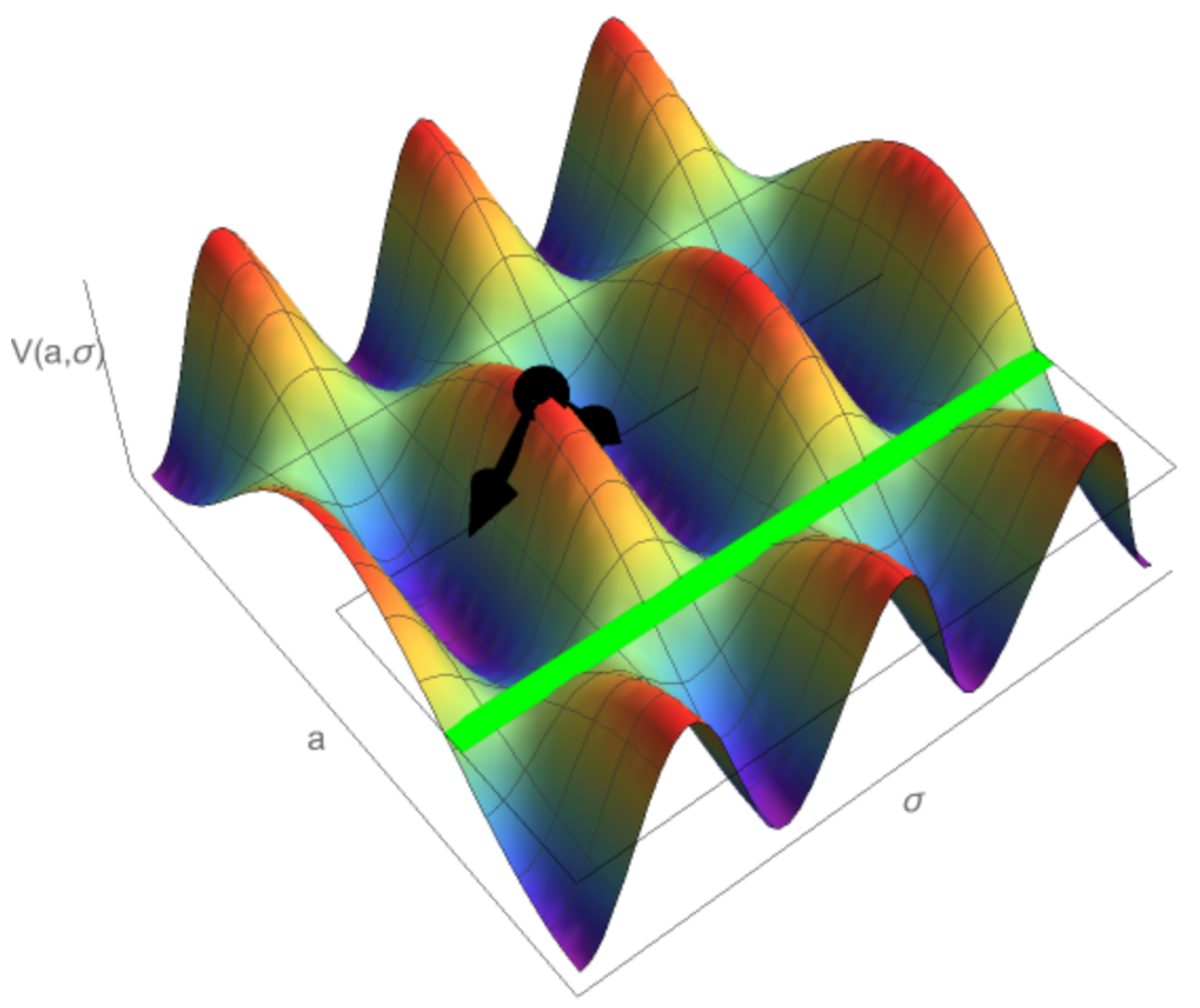}
\caption{ A typical plot of the QCD-axion potential $V(a,\sigma)$ in the $SU(N_c=2)$ case (but not to scale). Since $f\gg\Lambda$, the potential is very steep in the $\sigma$ direction. Thus, any small fluctuations near the black dot, which indicates the initial value of $a\approx \pi f$, will cause the inflation to proceed very quickly in the $\sigma$ direction (as indicated by the arrows), and thus, inflation ends abruptly. However, the axion can slowly roll down the potential provided that we start inflating near $a\approx \frac{\pi}{2}f$, irrespective of the initial value of $\sigma$, which is indicated by the thick green line.} 
\label{Axion potential su2}
\end{center}
\end{figure}

Inspection of Eq. (\ref{inflation su2}) reveals that slow-roll inflation cannot be sustained at values of $|a|\sim \pi f$ since in this case one finds $V(a,\sigma)\cong 2\Lambda^4\left(1+\cos\frac{\sigma}{\Lambda}\right)$. The latter form of the potential does not satisfy the slow-roll condition $\epsilon \equiv -\frac{\dot H}{H^2}\ll1$,  given that $\Lambda\ll M_P$,  a very sensible assumption in any reliable effective field theory. This behavior was confirmed numerically and is depicted in FIG. \ref{Axion potential su2}. The dynamics here is drastically different compared to  models of axion inflation, Eq. (\ref{axion Lagrangian}), where one starts near the hilltop at  $|a|\sim\pi f$. This should not come as a surprise given our lengthy discussion in Section \ref{Theory and formulation}: near the hilltop both the axion and hadronic degrees of freedom become intertwined, which is ultimately tied to the BCF anomaly, leading to a fast roll down. 

\begin{figure}[t]
\begin{center}
\includegraphics[width=87mm]{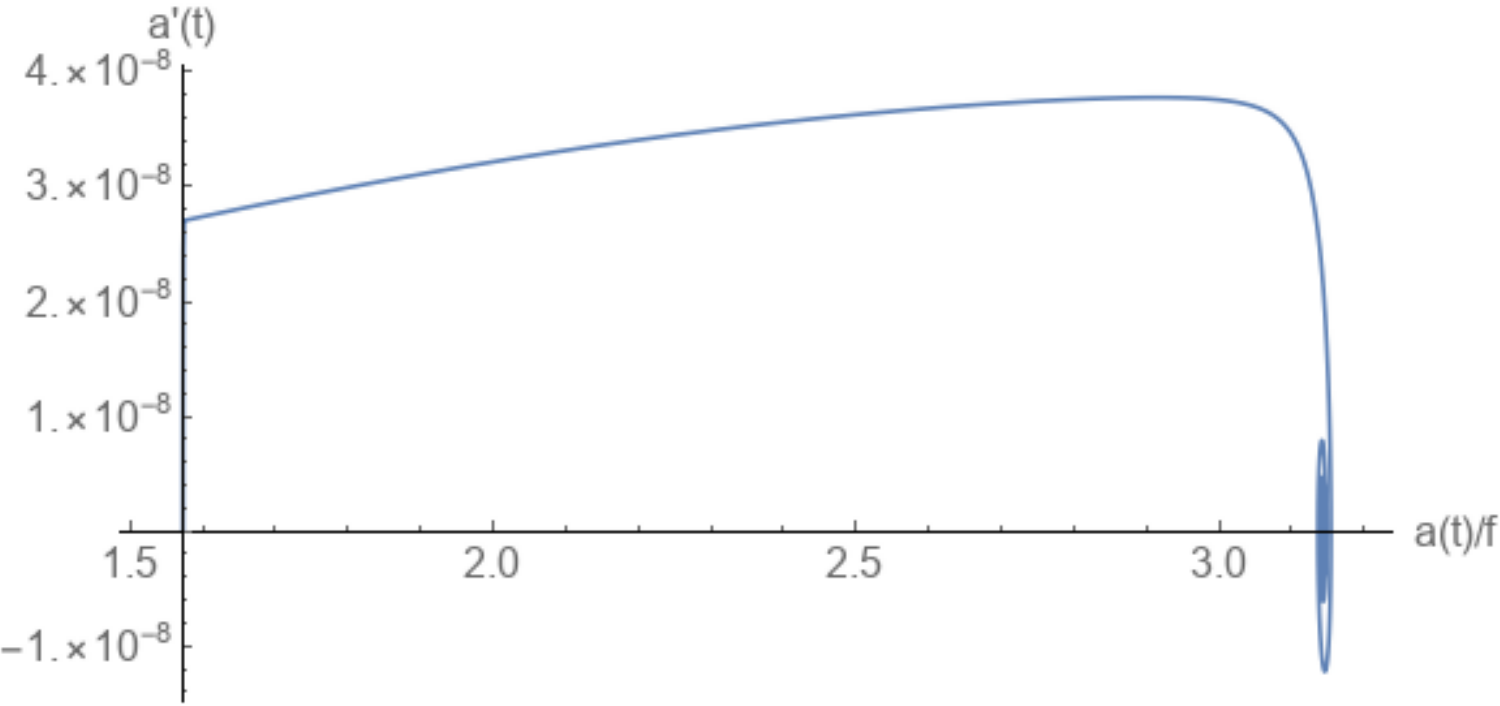}
\includegraphics[width=87mm]{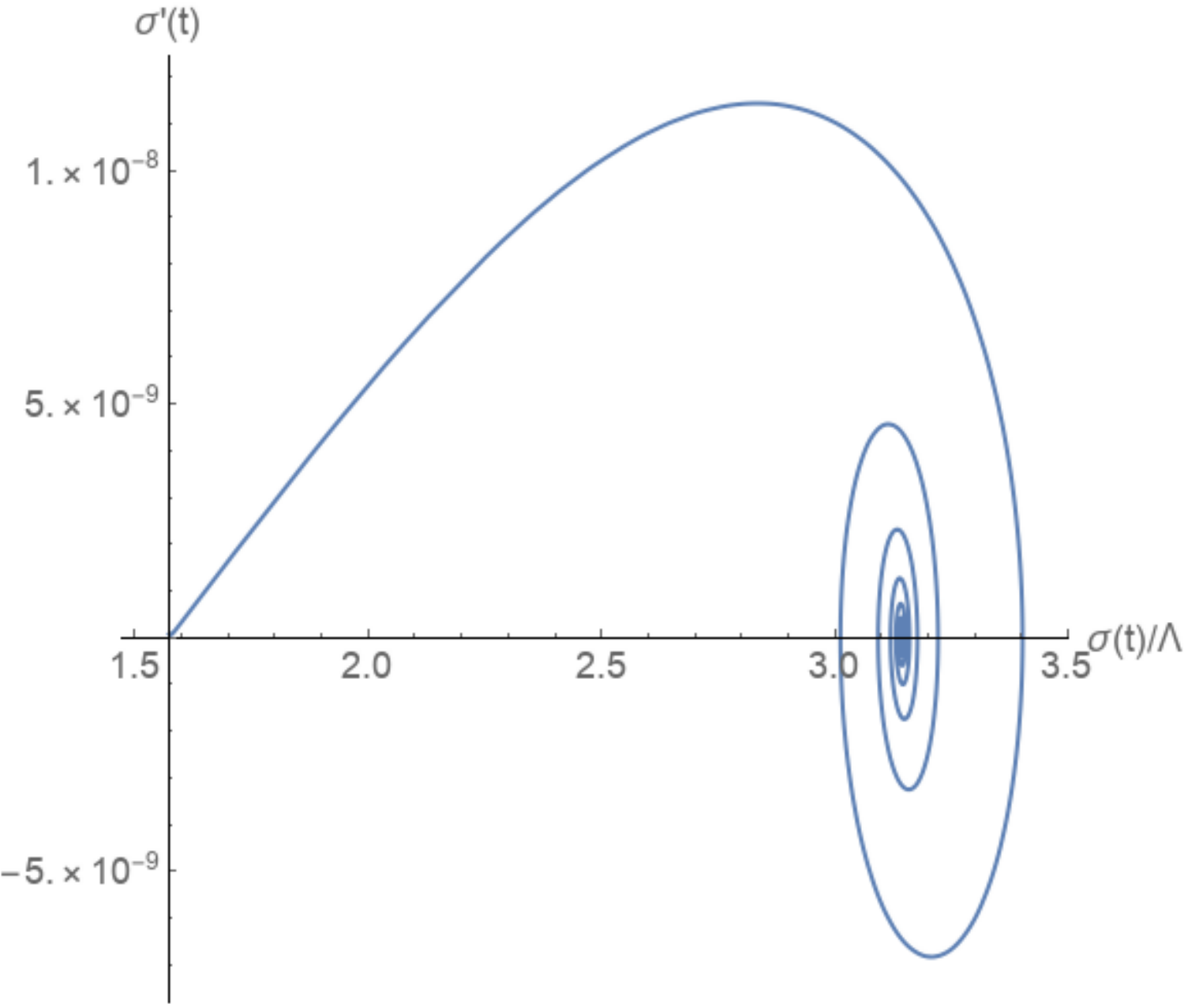}
\caption{The phase space of $a$ and $\sigma$. We take $\Lambda=10^{-3}\sqrt{8\pi} M_P$ and $f=6 \sqrt{8\pi} M_P$ and start inflation at $a\approx\frac{\pi}{2}f$ and $\sigma\approx \frac{\pi}{2} \Lambda$. We find, however, that the dynamics of inflation is insensitive to the initial values of $\sigma$. One can easily see that the axion velocity stays small throughout the inflation lifespan.}
\label{phase space su2}
\end{center}
\end{figure}

This, however, does not mean that slow-roll inflation is spoiled after taking the QCD effects into account. In fact, the potential (\ref{inflation su2}) can lead to a successful inflation, i.e., we can achieve $N_e\sim 50-60$, provided that we start inflating near $|a|\sim \frac{\pi}{2}f$, i.e., when the axion starts at the transition from the hill-like to valley-like regions of the potential, and  take $f> M_P$. This initial condition will restrict the motion of $\sigma$ between the high potential hills on either side, while the initial kick will cause $a$ to roll down from  $|a|\sim \frac{\pi}{2}f$ toward either $|a|\sim \frac{\pi}f$ or $a\sim 0$. Basically, avoiding the region where the QCD effects are important is mandatory in order to satisfy the slow-roll conditions. What is striking is that the QCD effects extend well beyond the QCD scale $\Lambda$ influencing the axion dynamics up to half-way through the axion field span, i.e., deep in the IR. This fascinating dynamics is a manifestation of the fact that 't Hooft anomalies cannot just disappear, they dictate the dynamics at all length scales.  The numerical investigation of this behavior is shown in FIG. \ref{phase space su2}. Unlike the case of traditional axion inflation, given by (\ref{axion Lagrangian}), here we find that there is no way one can obtain $50-60$ e-folds of inflation for values of $f<M_P$.  We conclude that the strong dynamics greatly influence the axion motion.

\begin{figure}[t]
\begin{center}
\includegraphics[width=87mm]{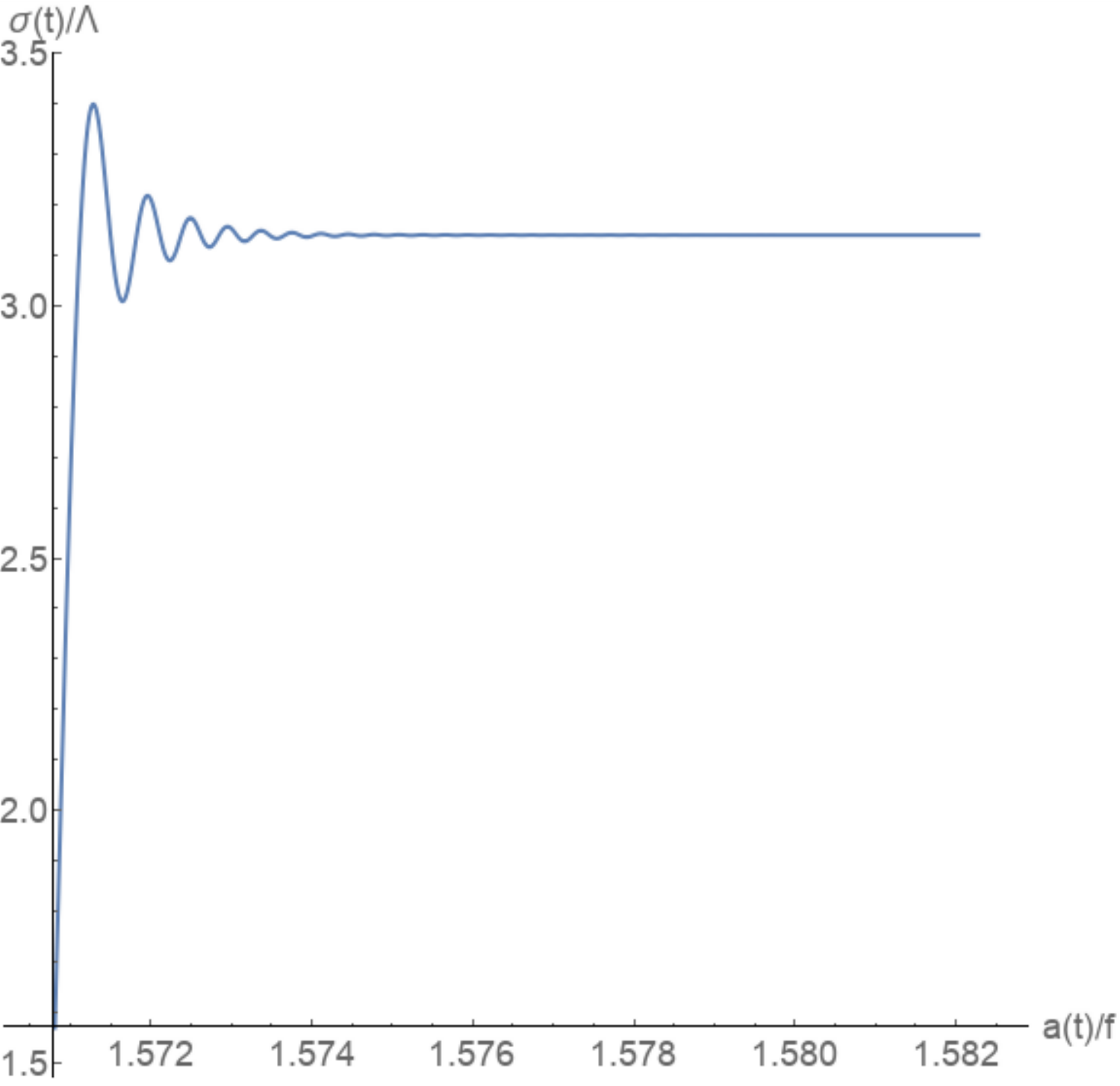}
\caption{The parametric relation between $\sigma$ and $a$. We take $\Lambda=10^{-3}\sqrt{8\pi} M_P$ and $f=6\sqrt{8\pi} M_P$. While the axion is rolling down from $\frac{\pi}{2}f$ to $\pi f$, the field $\sigma$ is almost instantaneously frozen (after $6$ e-folds in this example) at $\pi \Lambda$, which is consistent with the analytical finding Eq. (\ref{parametric sigma vs a}).}
\label{sigma vs a}
\end{center}
\end{figure}

\begin{figure}[t]
\begin{center}
\includegraphics[width=87mm]{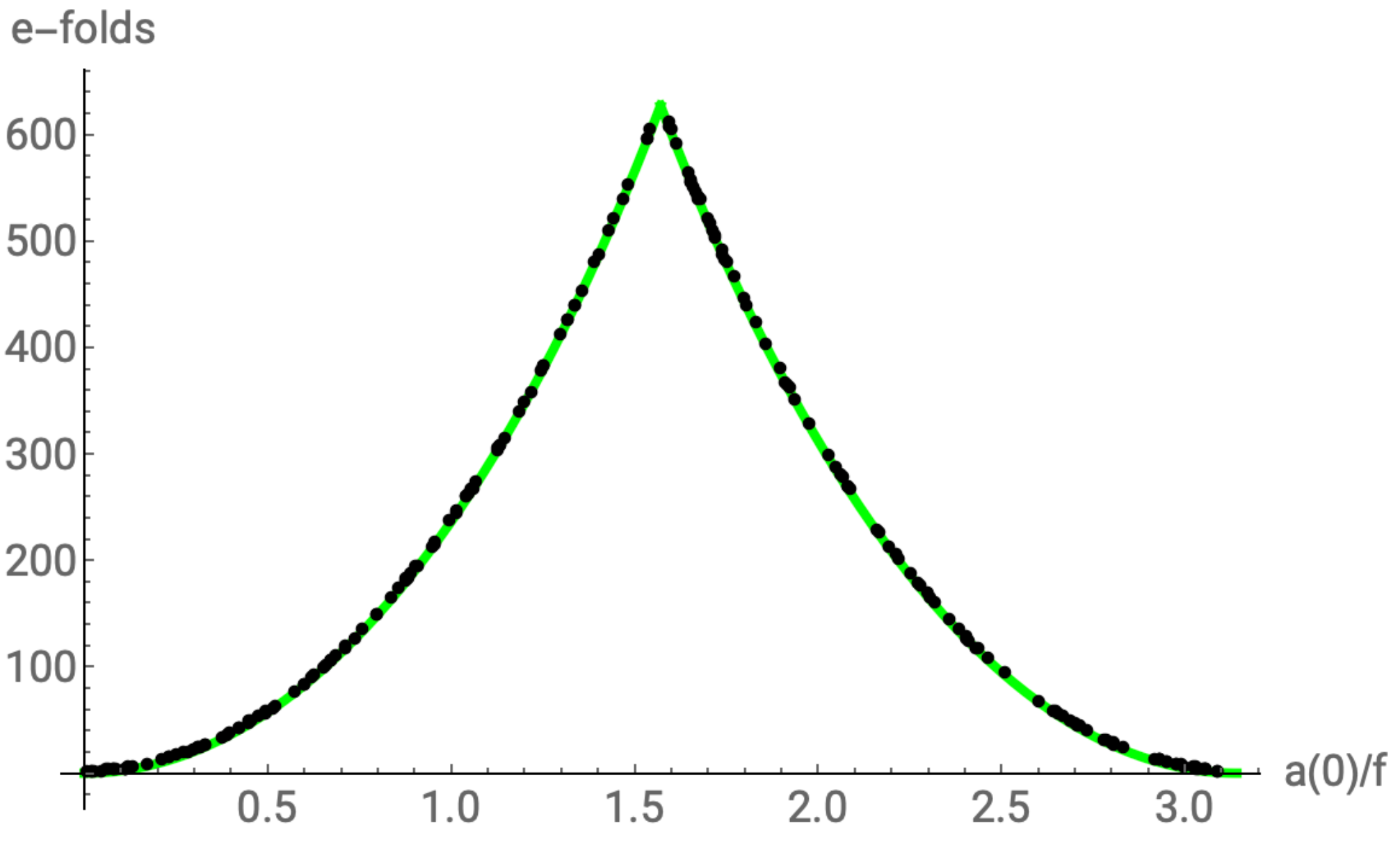}
\caption{The dependence of the number of e-folds on the initial value of $a$. The scattered black points are the numerical values, we take $\Lambda=10^{-3}\sqrt{8\pi} M_P$ and $f=6\sqrt{8\pi} M_P$, while the continuous green line is the analytical expression Eq. (\ref{N max}).  The maximum number is achieved by starting the inflation at $a(0)\approx\frac{\pi}{2}f$. We also checked that the number of e-folds is very insensitive to the value of $\Lambda$. It changes by less than 5\% as $\Lambda$ changes between $10^{-2}M_P$ and $10^{-5}M_P$.}
\label{efolds vs initial conditions}
\end{center}
\end{figure}

\begin{figure}[t]
\begin{center}
\includegraphics[width=87mm]{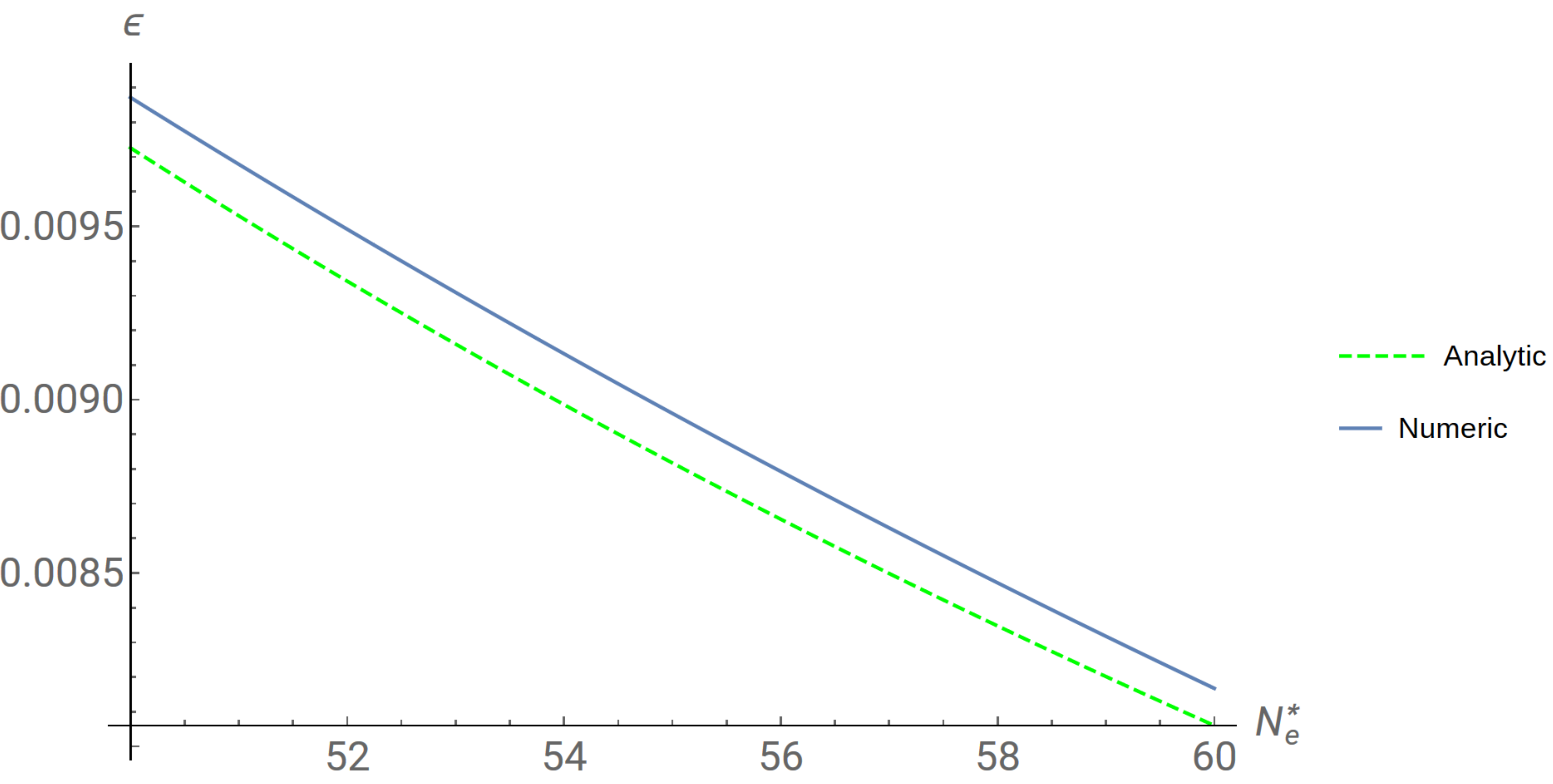}
\includegraphics[width=87mm]{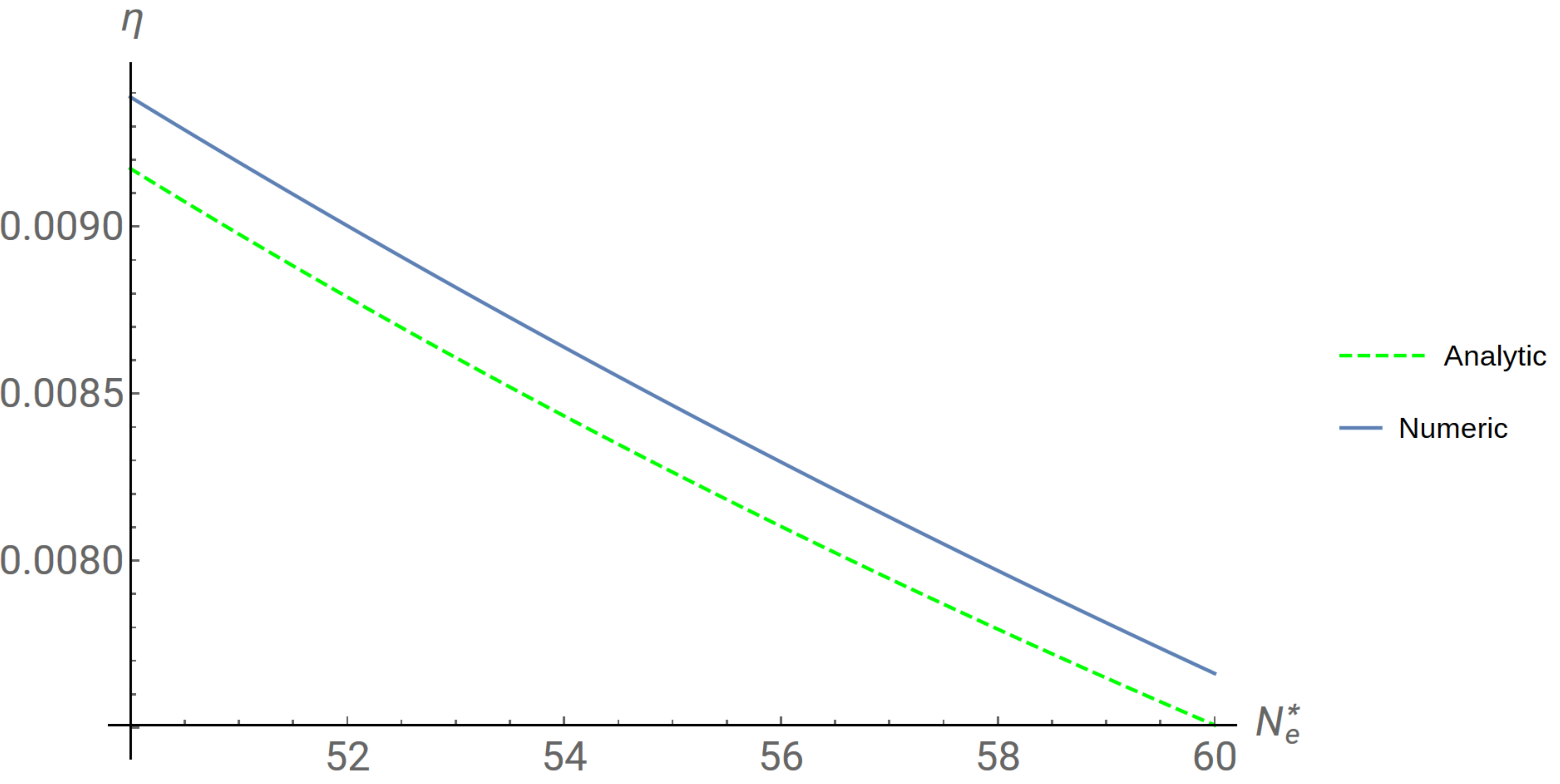}
\caption{The numerical values of $\epsilon$ and $\eta$ (the solid lines) against the analytical expressions (dotted lines) given in Eq. (\ref{analytical epsilon and eta}) as functions of $N_e^\star$, the number of e-folds remaining before the end of inflation. We take $\Lambda=10^{-3}\sqrt{8\pi} M_P$ and $f=6\sqrt{8\pi} M_P$. The discrepancy between the analytical and numerical solutions is less than $2\%$. We also find that changing the value of $\Lambda$ has a little effect on both $\epsilon$ and $\eta$, in agreement with Eq. (\ref{analytical epsilon and eta}) that does not depend explicitly on $\Lambda$. }
\label{numerical epsilon and eta}
\end{center}
\end{figure}

\begin{figure}[t]
\begin{center}
\includegraphics[width=87mm]{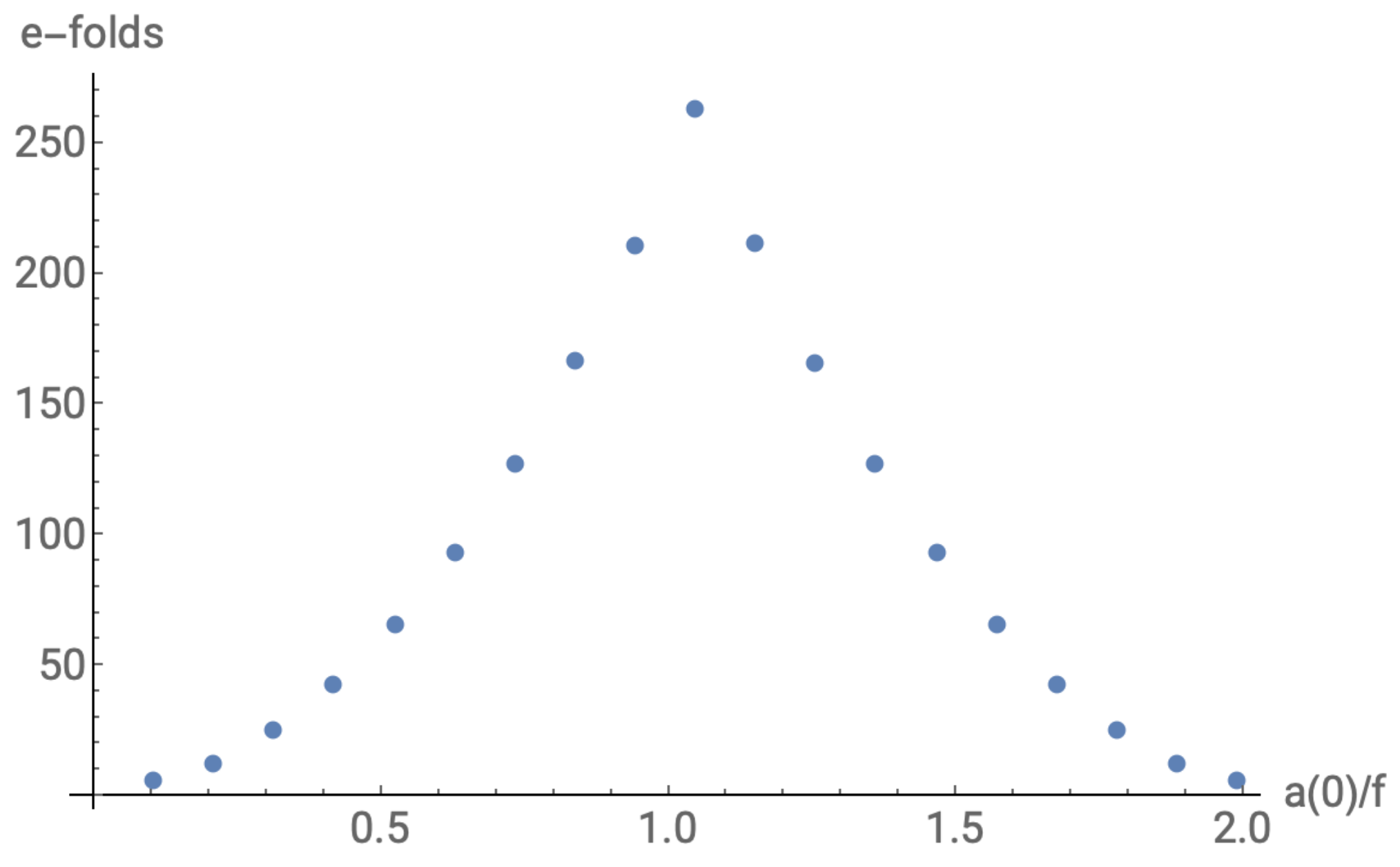}
\includegraphics[width=87mm]{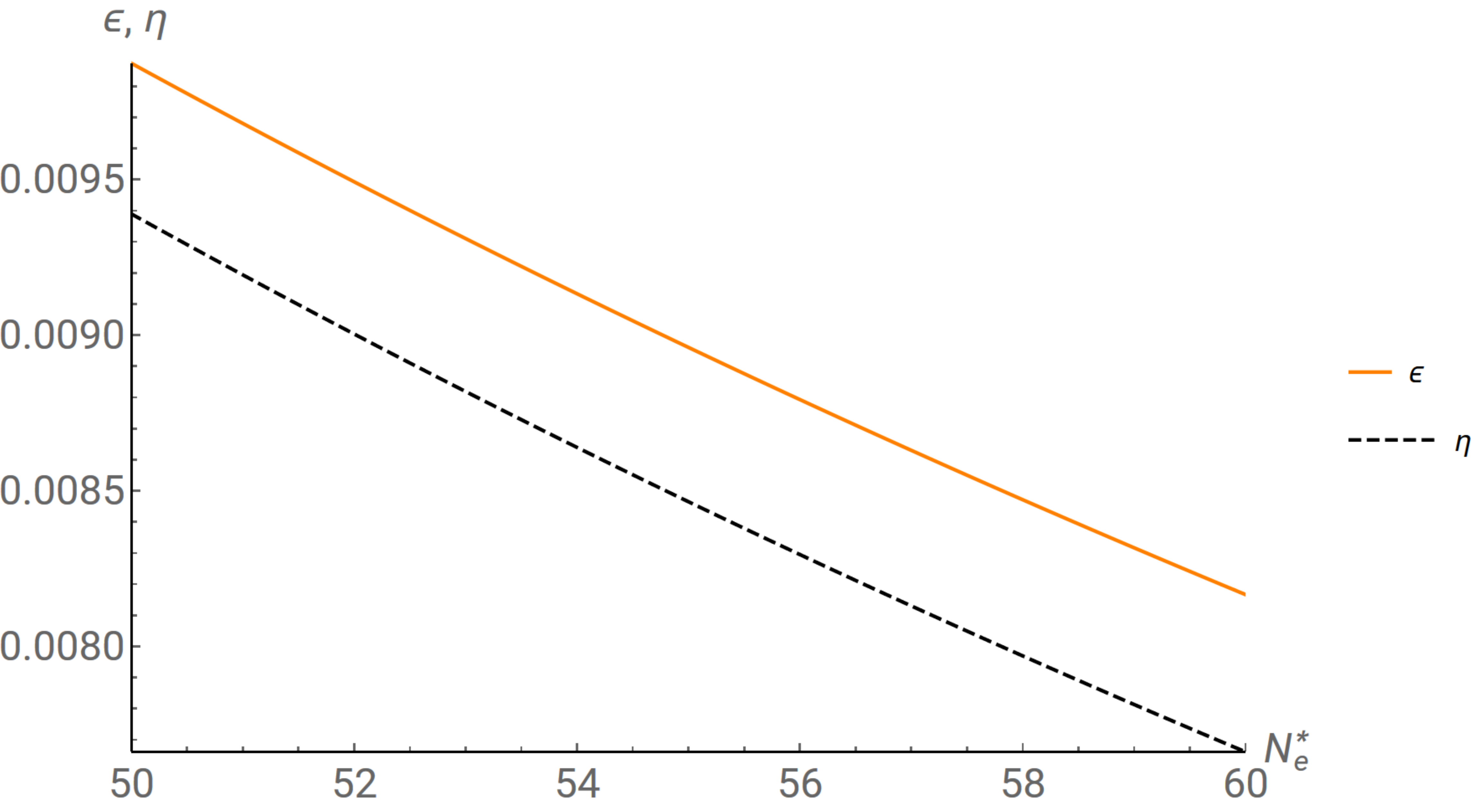}
\caption{The numerical data for the $SU(N_c=3)$ case. The  roots are $\bm \alpha_1=\left(\sqrt{2},0\right)$, $\bm \alpha_2=\left(-\frac{1}{\sqrt{2}},\sqrt{\frac{3}{2}}\right)$, $\bm\alpha_3=\left(-\frac{1}{\sqrt 2},-\sqrt\frac{3}{2}\right)$; see \cite{Georgi:1999wka}.  We take $f=6\sqrt{8\pi} M_P$ and $\Lambda=10^{-3}\sqrt{8\pi} M_P$. The top template shows the total number of e-folds as a function of the initial position of the axion. An initial position at $\frac{\pi}{3}f$ yields the largest value of the number of e-folds. The bottom template shows the values of $\epsilon$ and $\eta$ as functions of the number of e-folds remaining before the end of inflation. We can see that these values are almost identical to the ones in the $SU(N_c=2)$ case, and that the color group has a little effect on axion inflation once the axion is in the safe zone.}
\label{su3 data}
\end{center}
\end{figure}

We can also obtain analytical expressions of the slow-roll dynamics. Without loss of generality, and according to our numerical investigation, we can always restrict the axion motion in the interval $\frac{a}{f}\in [\frac{\pi}{2}, \pi]$, such that the axion starts rolling down near $\pi/2$. Also, we can assume that $\sigma$ starts near $\frac{\sigma^i}{\Lambda}=\frac{\pi}{2}$. We dub the interval near $\frac{a}{f}\cong \frac{\pi}{2}$ as the {\em safe zone}. We  numerically checked that the initial value of $\sigma$ doesn't affect the dynamics or any of the conclusions.  Neglecting the second derivative terms in the equations of motion (\ref{full system}) and integrating, we obtain:
\begin{eqnarray}
\cos \frac{\sigma}{\Lambda}=-\sqrt{1-\left(\sin\frac{a}{f}\right)^{\frac{2f^2}{\Lambda^2}}}\,,
\label{parametric sigma vs a}
\end{eqnarray}
and since $f\gg \Lambda$, we find that $\cos\frac{\sigma}{\Lambda}\approx-1$ is a very good approximation, which we use in the rest of analysis. This behavior is checked numerically and depicted in FIG.\ref{sigma vs a}.  As before, we can obtain the number of e-folds by performing the integral $\int_{a_i}^{a_f} da \frac{H}{\dot a}$ and using the Friedman's equation after neglecting $\dot a$ and $\dot \sigma$. Assuming that the inflation ends at $a_f\cong \pi f$, which is a very good approximation as we will see from the values of $\epsilon$ and $\eta$, we find 
\begin{eqnarray}
\sin \frac{a_i}{2f}=e^{-\frac{M_P^2N_e}{2 f^2}}\,,
\label{efolds QCD}
\end{eqnarray}
 and  the maximum number of e-folds is obtained by starting the inflation at $a=\frac{\pi}{2}f$:
\begin{eqnarray}
N_e^{\mbox{max}}=-\left[\frac{2 f^2}{M_P^2}\log\left( \sin \frac{a_i}{2f}\right)\right]_{a_i=\frac{\pi}{2}f}\,.
\label{N max}
\end{eqnarray}
For example, taking $f=6\sqrt{8\pi} M_P$ we find $N_e^{\mbox{max}}\cong 627$, which is in an excellent agreement with the numerical solution, see FIG. \ref{efolds vs initial conditions}. We also used our numerical scheme in order to check  that starting the inflation at values of $a$ slightly larger or smaller than $\frac{\pi}{2}f$ results in a smaller number of e-folds. Using Eq. (\ref{N max}) or our numerical code, we find that  $f \lesssim M_P$ gives $N_e \lesssim 0.7$, and thus, one cannot achieve inflation once QCD backreaction  is taken into account, in contradistinction with the single-field inflation. We also find that the minimum value of $f$ that  yields $60$ e-folds is $f\sim 9.4 M_P$.

If we assume that $a$ is randomly distributed between $0$ and $2\pi f$ from one horizon volume to another in a multiverse, then the probability of being in a region where inflation proceeds successfully, i.e., gives $N_e\sim 50-60$,  is given by $P=4\int_{\frac{\pi}{2 f}}^{\frac{a_{max}}{f}} \frac{d a}{2\pi f}$, where $a_{max}$ is the maximum value of $a_i$ in Eq. (\ref{efolds QCD}) that gives the minimum number of the required e-folds. Thus, we find 
\begin{eqnarray}
P=\frac{2}{\pi}\left[2\sin^{-1}\left(e^{-\frac{M_P^2 N_e}{2 f^2}}\right)-\frac{\pi}{2}\right]\,.
\end{eqnarray}
For example, using $f=6\sqrt{8\pi} M_P$ and $f=50 \sqrt{8\pi} M_P$ gives $P=67\%$ and $P=96\%$, respectively,  for $N_e=60$.

We can proceed to calculate the slow-roll parameters: using the equations of motion (after neglecting the second derivatives), the Friedmann's equation (after neglecting the first derivatives), and the approximation $\cos\frac{\sigma}{\Lambda}\approx-1$,  we find\footnote{We would like to emphasize that there is no heterogeneity between the tan formula that appears here and cotan formula that appears near the discussion of the traditional axion inflation. The formula in Eq. (31) is derived assuming that we start inflating near $a/f=\pi/2$, while the cotan formula  is derived  assuming that we start inflating near $a/f=\pi$. Notice that this is the main difference between the traditional axion inflation (where we assume that we start inflating near the hilltop at $a/f=\pi$) and the inflation model we are studying in this work, which takes the QCD effects into account. As we emphasize, in the latter case one cannot inflate near $a/f=\pi$ since this initial condition spoils the slow-roll parameters. So in summary, the tan versus cotan formula reflects the fact that we start inflating at different points in the field space that are shifted by $\pi/2$.}
\begin{eqnarray}
\epsilon= \frac{M_P^2}{2 f^2}\tan^2\frac{a}{2f}\,,\quad
\eta=-\frac{M_P^2}{f^2}\frac{\cos \frac{a}{f}}{1+\cos\frac{a}{f}}\,.
\end{eqnarray}
It can be easily checked that these parameters stay small during inflation provided that $f>9M_P$. We can also express $\epsilon$ and $\eta$ in terms of $N_e^\star$, the number of e-folds remaining before the end of inflation:
\begin{eqnarray}
\nonumber
\epsilon(N_e^\star)&=&\frac{M_P^2}{2 f^2}\frac{1}{e^{\frac{M_P^2 N^\star_e}{f^2}}-1}\,,\\
\eta(N_e^\star)&=&\frac{M_P^2}{2 f^2}\frac{2-e^{\frac{M_P^2 N^\star_e}{f^2}}}{e^{\frac{M_P^2 N^\star_e}{f^2}}-1}\,.
\label{analytical epsilon and eta}
\end{eqnarray}
Interestingly, these are the exact same expressions we obtained above before taking the strong dynamics effects into account. This result is not unexpected since the $\sigma$ field becomes frozen near $\pi \Lambda$, and effectively the dynamics is governed by a single field $a$. The caveat, however, is that this is true only for values of $a/f$ in the safe zone near $\frac{\pi}{2}$. Otherwise, the QCD effects become vicious and completely spoil inflation.    As we shall show below, this observation will play a pivotal role in conclusions about the power spectrum and tensor perturbations. In FIG. \ref{numerical epsilon and eta} we compare the analytical results (\ref{analytical epsilon and eta}) with numerical calculations to find an excellent agreement, with difference  less than $2\%$.  

Before we conclude this section, we also present the numerical solution of the $SU(N_c=3)$ case, see FIG. \ref{su3 data}. In order to be able to compare with  $SU(N_c=2)$ we make the substitution $f\rightarrow 3f/N_f T_{\cal R}$ in Eq. (\ref{model of axion inflation 1}).  The maximum number of e-folds will be achieved if we start inflating at $a\approx \frac{2\pi k}{3}f$, $k=1,2$, while the initial values of $\bm \sigma$ have almost no effect on the dynamics, exactly as in $SU(2)$.  We also compute the slow-roll parameters $\epsilon$ and $\eta$ at $N_e^\star\sim 50-60$ e-folds before the end of inflation to find almost an exact match with the $SU(2)$ case. This behavior indicates that the rank of the gauge group has a minor effect on the axion (at least for small values of $N_c$) and that strong dynamics will cease to affect inflation once the axion is rolling down in the safe zone.

\subsection{Quantum fluctuations: the QCD effects} Now, it is time to ask about the effect of QCD on the quantum fluctuations during inflation, and hence, on the CMB power spectrum. We limit our treatment to the $SU(N_c=2)$ case. To this end, we proceed as usual by writing both $a$ and $\sigma$ fields as classical backgrounds and small perturbations: $a=a_c+\delta a$, $\sigma=\sigma_c+\delta \sigma$. Then, we substitute in the equations of motion (after restoring the dependence on the spatial coordinates $\nabla^2 a$ and $\nabla^2\sigma$) to obtain:
\begin{eqnarray}
\nonumber
\delta \ddot a+3H\delta \dot a +\frac{\partial ^2 V}{\partial a^2}\delta a +\frac{\partial^2 V }{\partial a \partial \sigma}\delta \sigma +\frac{k^2}{b^2}\delta a=0\,,\\
\delta \ddot \sigma+3H\delta \dot \sigma+ \frac{\partial^2 V }{\partial a \partial \sigma}\delta a+\frac{\partial ^2 V}{\partial \sigma^2}\delta \sigma+\frac{k^2}{b^2}\delta \sigma=0\,,
\end{eqnarray}
where $b$ is the scale factor and $k$ is the comoving wave number. Further, we define $\delta \chi_1\equiv b \delta a$, $\delta \chi_2\equiv b \delta \sigma$, make the change of variables from the cosmic time $t$ to conformal time $\tau$ via $d\tau=dt/b$, and use the approximation $b(\tau)=-\frac{1}{H\tau (1-\epsilon)}$. In this approximation we assume that the Hubble parameter stays constant over the course of inflation, which is a very good approximation in the case of a single-field inflation. As we discussed above, one can completely forget about the classical dynamics of $\sigma$ during the full span of inflation (when the axion is in the safe zone) since $\sigma$ spends most of its life near $\sigma\approx \pi \Lambda$. Effectively, we have a single field inflation and the approximation $b(\tau)=-\frac{1}{H\tau (1-\epsilon)}$ is justified. The purpose of the present analysis is to check whether the fluctuations, rather than the classical dynamics, of $\sigma$ become important during any stage of inflation. 

After a straight forward calculation we obtain:
\begin{eqnarray}
\nonumber
\left[\frac{d^2 }{d\tau^2}+k^2+\frac{-2-3\epsilon+3\eta_{aa}}{\tau^2} \right]\delta\chi_1+\frac{3\eta_{a\sigma}}{\tau^2}\delta\chi_2=0\,,\\
\nonumber
\left[\frac{d^2 }{d\tau^2}+k^2+\frac{-2-3\epsilon+3\eta_{\sigma\sigma}}{\tau^2} \right]\delta\chi_2+\frac{3\eta_{a\sigma}}{\tau^2}\delta\chi_1=0\,,\\
\end{eqnarray}
where we defined:
\begin{eqnarray}
\eta=\eta_{aa}\equiv \frac{\frac{\partial^2 V}{\partial a^2}}{3H^2}\,, \eta_{\sigma\sigma}\equiv \frac{\frac{\partial^2 V}{\partial \sigma^2}}{3H^2}\,, \eta_{a\sigma}\equiv \frac{\frac{\partial^2 V}{\partial a\partial\sigma}}{3H^2}\,.
\end{eqnarray}
At this stage we  use the results of the previous section to write $\eta_{aa}$, $\eta_{\sigma\sigma}$, and $\eta_{a\sigma}$ as functions of $N_e^\star$: 
\begin{eqnarray}
\nonumber
\eta(N_e^\star)&=&\eta_{aa}(N_e^\star)=\frac{M_P^2}{2 f^2}\frac{2-e^{\frac{M_P^2 N^\star_e}{f^2}}}{e^{\frac{M_P^2 N^\star_e}{ f^2}}-1}\,,\\
\nonumber
\eta_{\sigma\sigma}(N_e^\star)&=&\frac{M_P^2}{2 \Lambda^2}\frac{2-e^{\frac{M_P^2 N^\star_e}{ f^2}}}{e^{\frac{M_P^2 N^\star_e}{ f^2}}-1}\,,
\eta_{a\sigma}(N_e^\star)\approx0\,.\\
\end{eqnarray}
Since $\eta_{a\sigma}(N_e^\star)\approx0$, the fluctuations $\delta \chi_1$ and $\delta \chi_2$ decouple. Moreover, since $|\eta(N_e^\star)|\ll|\eta_{\sigma\sigma}(N_e^\star)|$, the fluctuations of the $\sigma$ field stay in the vacuum as the $a$ fluctuations exit the horizon\footnote{Notice that $\eta$ is the ratio of the fluctuation mass to the Hubble parameter. Therefore having $|\eta|\gg 1$ means that the amplification of the fluctuations is highly suppressed.}. This is expected since  $\sigma$ is orders of magnitude heavier than $a$; remember that $\Lambda\ll f$. Thus,  the CMB power spectrum and tensor perturbations are solely governed by the dynamics and fluctuations of the axion and that the strong dynamics has a negligible effect on both quantities. The caveat, however, is that this is only true as long as we start inflation in the safe zone near $a\cong \frac{\pi}{2}f$. 

We conclude that the spectral tilt and the tensor to scalar ratio are given by the single-field expressions (\ref{tilt and r}), and thus, strong dynamics does not alter these quantities as long as the axion rolls down in the safe zone.

\section{Final Comments}
\label{Final Comments}

In this letter we investigated the role played by the strong dynamics in models of natural inflation. We argued that the single-field effective potential breaks down as the axion makes large-field excursions and that it overlooks the effects of the strong dynamics at the hilltop. We also argued that the intertwining between the axion and hadrons near the hilltop is ultimately tied to an 't Hooft anomaly. 

Since $\Lambda\gg m_a\sim\frac{\Lambda^2}{f}$, one   would naively expect that hadrons  have a negligible effect on the axion dynamics, unless we start extremely close to the hilltop within a narrow strip of width $\sim \frac{m_a}{\Lambda}\sim 10^{-4}$  around $\pi$. Interestingly, the influence of the hadrons on axions occurs not only in a small region, it actually extends further away changing the conclusions in a dramatic way. For example, we found that using $f<M_P$ does not yield the e-folds required to solve the problems of the standard Big-Bang Cosmology. Yet, taking $f> 9M_P$, we were also able to identify a safe zone where the hadrons decouple. We found that as the axion rolls-down the safe zone, the power spectrum and tensor to scalar ratios are identical to the values obtained from the single-field potential. Such conclusions are  independent of the number of colors, and we also expect them to be generic in other  QCD-axion inflationary models.

Finally, we point out that our analysis raises a  question about the axion misalignment mechanism \cite{Turner:1985si}. In this scenario the misalignment  can produce the observed dark matter (DM) abundance if $a/f$ is taken sufficiently close to $\pi$. The idea is that in order to generate the observed DM abundance,  the axion should be frozen at the hilltop until the era of QCD phase transition. As the Hubble parameter becomes comparable to the strong scale,  the axion starts oscillating and  produce DM. One expects, however, the effects of strong dynamics to be enhanced near $a/f\cong \pi$, which can affect the DM abundance. Since we are considering QCD at temperatures comparable to $\Lambda$, it is not yet clear how one can take the strong dynamics into account and whether the BCF anomaly survives at finite temperatures. A detailed study of this problem is left for the future. 

\acknowledgements
We would like to thank Erich Poppitz for raising the question about axion inflation in the light of the BCF anomaly, for several discussions, and comments on the manuscript. Also, we would like to thank Lorenzo Sorbo for comments on the manuscript. This work is supported by NSF grant PHY-2013827. 

\bibliographystyle{apsrev4-1}
\bibliography{references}

\end{document}